\documentclass[twocolumn]{aastex631}

\usepackage{wrapfig}
\usepackage{multirow}
\usepackage{amsmath}

\begin{document}

\title{Rise of the Forsaken Relics: Connecting Present-day Stellar Streams and Phase-mixed galaxies \\ to the Epoch of Reionization}

\correspondingauthor{Aritra Kundu}
\email{aritra@sas.upenn.edu}

\author[0000-0001-8746-4753]{Aritra Kundu}
\affiliation{Department of Physics \& Astronomy, University of Pennsylvania, 209 S 33rd St., Philadelphia, PA 19104, USA}

\author[0000-0003-3939-3297]{Robyn Sanderson}
\affiliation{Department of Physics \& Astronomy, University of Pennsylvania, 209 S 33rd St., Philadelphia, PA 19104, USA}

\author[0000-0002-3950-9598]{Adam Lidz}
\affiliation{Department of Physics \& Astronomy, University of Pennsylvania, 209 S 33rd St., Philadelphia, PA 19104, USA}

\author[0000-0003-0965-605X]{Pratik J. Gandhi}
\affiliation{Department of Astronomy, Yale University, P.O. Box 208101
New Haven, CT 06520-8101 USA}

\author[0000-0003-0603-8942]{Andrew Wetzel}
\affiliation{Department of Physics \& Astronomy, University of California, Davis, One Shields Avenue, Davis, CA 95616, USA}

\author[0000-0002-1109-1919]{Robert Feldmann}
\affiliation{Department of Astrophysics, Universit\"at Z\"urich, Winterthurerstrasse 190, Zurich, 8057, Switzerland}

\author[0000-0001-5214-8822]{Nondh Panithanpaisal}
\affiliation{The Observatories of the Carnegie Institution for Science, 813 Santa Barbara St, Pasadena, CA 91101, USA}
\affiliation{TAPIR, California Institute of Technology, Pasadena, CA 91125, USA}

\author[0009-0008-8901-2206]{Jasjeev Singh}
\affiliation{Department of Physics \& Astronomy, University of Pennsylvania, 209 S 33rd St., Philadelphia, PA 19104, USA}

\author[0000-0002-9604-343X]{Michael Boylan-Kolchin}
\affiliation{Cosmic Frontier Center, Department of Astronomy, The University of Texas at Austin, \\2515 Speedway, Austin, TX 78712, USA}









\begin{abstract}
The ``near-far" approach to studying reionization leverages the star formation histories of the Milky Way (MW) or Local Group (LG) galaxies, derived from resolved photometry, to infer the low-mass/faint end of the stellar mass functions (SMFs) or the ultraviolet luminosity functions (UVLFs) of high-redshift galaxies ($z \gtrsim 6$), beyond the current JWST detection limits ($M_{\mathrm{UV}} \gtrsim -15$).
Previous works considered only intact low-mass galaxies in the MW and LG, neglecting disrupted galaxies such as stellar streams and phase-mixed objects. 
Using the FIRE-2 simulations, we show that these disrupted galaxies contribute up to $\sim$50\% of the total stellar mass budget of the proto-MW/LG at $z =6-9$. 
Including all the progenitors of these disrupted galaxies improves the normalization of the recovered SMFs/UVLFs by factors of $\sim$2--3 and reduces the halo-to-halo variation in the slope by $\sim$20--40\%. 
This enables robust constraints down to at least the resolution limit of the simulations, near $M_\star \sim 10^{5} M_\odot$ or $M_{\mathrm{UV}} \sim -10$ at $z \gtrsim 6$. 
We also show that ``fossil record" reconstructions---which assume each present-day system descends from a single reionization-era progenitor---are sensitive to the stellar-mass/UV-magnitude thresholds, which introduces bias in the inferred slopes at the low-mass/faint end. 
Additionally, we demonstrate that neglecting disrupted systems underestimates the contribution of galaxies with $M_{\mathrm{UV}} \lesssim -15$ to the reionization-era UV luminosity density. 
Finally, we estimate that a significant fraction ($\sim50\%$) of streams with $M_\star \gtrsim 10^{6} M_\odot$ at $z$=0 should be detectable from upcoming Rubin Observatory and Roman Space Telescope observations.
\end{abstract}



\section{Introduction} \label{sec:intro}

The Epoch of Reionization (EoR) is the time period when the first stars, galaxies, and accreting black holes emitted ultraviolet (UV) radiation and gradually photo-ionized neutral hydrogen in the intergalactic medium (IGM) throughout the universe \citep{Barkana_Loeb_2001, Furlanetto_et_al_2006}. The timing and nature of this process remain highly uncertain, while the properties of the ionizing sources are currently unknown. The most likely scenario is that the UV radiation from star-forming galaxies drives reionization, with the process completing around $z \sim 6$ \citep{Shapiro_etal_1994, Madau_etal_1999, Faucher-Giguere_etal_2008b, Becker_and_Bolton_2013}, although precisely which galaxies dominate is unclear. The Hubble Space Telescope (HST) has directly observed faint reionization-era galaxies down to $M_{\mathrm{UV}} \sim -13$ in fields behind lensing clusters \citep{Finkelstein_etal_2015, Bouwens_etal_2017, Livermore_etal_2017, Atek_etal_2018, Bouwens_etal_2022a}, albeit with sizable uncertainties at the faint end. Ongoing and future observations with the James Webb Space Telescope (JWST) are pushing these measurements to $z \gtrsim 10$ and slightly fainter magnitudes (e.g., $M_{\mathrm{UV}} \sim -12$) \citep{Finkelstein_etal_2023, Leung_etal_2023, Navarro-Carrera_etal_2023, Perez-Gonzalez_etal_2023, Atek_etal_2024, Finkelstein_etal_2024, Adams_etal_2024, Harikane_etal_2025, Chemerynska_etal_2025, Atek_etal_2025}. 

However, some reionization models predict that fainter yet more numerous low-mass galaxies are major contributors to the total budget of ionizing photons, potentially with $M_{\mathrm{UV}}$ down to $ \sim -8$ \citep{Kuhlen_and_Faucher-Giguere_2012, Wise_etal_2014, Robertson_etal_2015, Atek_etal_2015, Ma_etal_2018a, Ma_etal_2018b, Finkelstein_etal_2019, Atek_etal_2024, Mascia_etal_2024, Simmonds_etal_2024, Harshan_etal_2024, Wu_and_Kravtsov_2024}.
Many of these sources will remain beyond the reach of even JWST \citep{Boylan-Kolchin_etal_2015, Boylan-Kolchin_etal_2016, Weisz_and_Boylan-Kolchin_2017}. While observations of lensed galaxies can probe fainter magnitudes (down to $M_{\mathrm{UV}} \sim -12$), they introduce significant systematic uncertainties due to magnification effects \citep{Atek_etal_2018, Bouwens_etal_2022b}. Additionally, low-mass galaxies provide valuable insights into the impact of stellar feedback and the nature of dark matter on galaxy formation, particularly at high redshifts \citep{Sun_and_Furlanetto_2016, Bullock_and_Boylan-Kolchin_2017, Sales_etal_2022, Sipple_and_Lidz_2024, Sipple_etal_2025}. Novel approaches for probing the faint/low-mass end of the $z \gtrsim 6$ galaxy UV luminosity function (UVLF)/stellar mass function (SMF) are hence valuable not only in determining the role of low-luminosity galaxies in reionizing the universe, but also in understanding feedback processes and the properties of DM.

One promising technique is the ``near-far" approach, which relies on studies of Milky Way (MW) satellite galaxies and Local Group (LG) low-mass galaxies at the present-day ($z = 0$) to infer properties of their $z \gtrsim 6$ progenitors \citep{Boylan-Kolchin_etal_2015, Boylan-Kolchin_etal_2016, Weisz_and_Boylan-Kolchin_2017, Weisz_etal_2014b}. This employs color-magnitude diagrams (CMDs) of nearby MW/LG low-mass galaxies from resolved HST/JWST photometry to reconstruct their star-formation histories (SFHs; \citealt{Weisz_etal_2014a, Weisz_etal_2014b}) and UV luminosities at $z \gtrsim 6$ \citep{Weisz_etal_2014c, Weisz_and_Boylan-Kolchin_2017}, using stellar population synthesis models. This indirect approach probes fainter magnitudes than direct high-redshift observations, thereby providing information about the faint-end/low-mass-end of the UVLFs/SMFs at $z \gtrsim 6$. 

Previous works applied this technique to intact low-mass galaxies in the LG \citep{Boylan-Kolchin_etal_2016, Weisz_and_Boylan-Kolchin_2017, Weisz_etal_2014c, Skillman_etal_2017, Savino_etal_2023, McQuinn_etal_2024}. This strategy assumes a one-to-one mapping from high-$z$ low-mass galaxies to their present-day LG counterparts, ignoring the cosmological assembly histories of the low-mass galaxies and, in the case of satellite galaxies, their host halos. One consequence of this assumption is that the high-$z$ SMF/UVLF inferred from the LG fossil record recovers only $\approx$15-20\% of the total amplitude \citep{Gandhi_etal_2024}. This is because it lacks information about the different progenitors that merged to form the present-day system. \citet{Gandhi_etal_2024} analyzed the impact of these mergers by including multiple progenitors of present-day low-mass galaxies in cosmological baryonic zoom-in simulations, and tested the near-far technique in terms of recovering the slope of the low-mass-end of the SMF at $z \gtrsim 6$. They found that the fossil record of surviving low-mass galaxies in the MW/LG can recover the slope at the faint/low-mass end of the UVLF/SMF of the proto-MW/LG at $z \gtrsim 6$, albeit with a large halo-to-halo scatter. 

Although these studies illustrate the general promise of the near-far technique, they do not yet fully account for the fact that most high-$z$ progenitors fail to survive intact as present-day MW/LG low-mass galaxies \citep{Santistevan_etal_2020, Horta_etal_2024}. In fact, most high-$z$ progenitor galaxies will be completely disrupted by mergers and tidal forces ($\sim50\%$ of the proto-MW/LG, \citealt{Gandhi_etal_2024}), leaving behind both coherent stellar streams and galaxy remnants that are now part of the equilibrium phase-space distribution defined by the host galaxy (``phase-mixed galaxies"). These merged substructures have systematically different properties from the intact satellites, such as lower metallicities due to earlier star formation \cite[e.g.,][]{Panithanpaisal_etal_2021, Cunningham_etal_2022, Horta_etal_2023}. The near-far technique usually ignores these disrupted galaxies when inferring the high-$z$ UVLF/SMF. However, the high-$z$ progenitors of these substructures also contribute to the ionizing photon budget at $z \gtrsim 6$, and their exclusion may lead to inaccurate conclusions regarding the role of low-luminosity galaxies in reionizing the universe. The reconstructed SMFs/UVLFs from intact galaxies alone require large corrections to the normalizations of these functions. 
Furthermore, simulations show a large scatter in the shape of the reconstructed SMFs/UVLFs across different MW/LG-like halos (halo-to-halo scatter) \citep{Gandhi_etal_2024}. Incorporating disrupted galaxies into the analysis could potentially help reduce this scatter and provide more robust constraints on the shape and normalization of reconstructed high-$z$ MW/LG SMF/UVLF. Hence, a comprehensive study of the impact of including disrupted galaxies on the near-far technique is crucial. 

Concurrent with more comprehensive theoretical tests of the near-far technique, it is important to assess the detectability of the disrupted galaxies for use in this approach. To date, the \textit{Gaia} mission and the Dark Energy Survey (DES) have detected significant numbers of low surface brightness ($\mu > 27$ mag/arcsec$^2$, \citealt{Johnston_etal_2008}) galaxies, such as stellar streams and phase-mixed galaxies, in the MW, greatly enhancing our understanding of the Milky Way's formation history \citep[e.g.,][]{Ibata_etal_2001, Newberg_etal_2002, Balbinot_etal_2016, Shipp_etal_2018, Necib_etal_2020, Belokurov_etal_2018, Martin_etal_2022}. However, the rich accretion history probed by the disrupted galaxies extends beyond $> 30$ $\mathrm{mag}$ $\mathrm{arcsec^{-2}}$ \citep{Laine_etal_2018}. Upcoming telescopes like the Vera C. Rubin Observatory \citep{Ivezic_etal_2019}, Roman Space Telescope \citep{Spergel_etal_2013}, and ARRAKIHS \citep{Guzman_etal_2022} will enable detections of these fainter low surface brightness substructures both within the MW and beyond \citep{Pearson_etal_2022a, Pearson_etal_2022b, Dey_etal_2023, Aganze_etal_2024}. This presents us with an opportunity to utilize these substructures to infer the properties of their high-$z$ progenitors and learn about their likely contribution to reionization, thereby also comprehensively testing the feasibility of the near-far technique in probing the EoR.

In this paper, we investigate these issues using a suite of simulated MW/LG-like galaxies from zoomed cosmological hydrodynamic simulations. 
In Section \ref{sec:simulations}, we describe the simulations used in this work. Section \ref{sec:substuctures_and_progs} discusses our criteria for classifying present-day substructures and our methodology for tracking them back to their simulated high-$z$ progenitors. Through Sections \ref{sec:mass_budget_and_struct_form}--\ref{sec:detectability}, we illustrate the impact of adding disrupted galaxies to the near-far technique and also test the feasibility of using them by imposing detectability constrains. Finally, we summarize these results in Section \ref{sec:summary}.

\section{Simulations} \label{sec:simulations}

\subsection{FIRE-2 Simulations} \label{subsec:fire2-sims}

We use cosmological zoom-in baryonic simulations from the Feedback in Realistic Environments (FIRE-2) project \citep{Wetzel_etal_2023, Wetzel_etal_2025} \footnote{\url{https://fire.northwestern.edu/}}. They are run with the \textsc{Gizmo} code \citep{Hopkins_2015, Hopkins_etal_2018a}, which uses a mesh-free, finite mass (MFM) 
Lagrangian method for accurate hydrodynamics. Baryonic physical processes like star formation and stellar feedback are implemented using the FIRE-2 numerical prescriptions \citep{Hopkins_etal_2018a}. The multiphase interstellar medium (ISM) in each simulated galaxy is captured with realistic radiative heating and cooling processes across temperatures from 10 K to 10$^{10}$ K. 

Simulated star particles form out of gas parcels satisfying a Jeans instability criterion, and inherit the mass and elemental abundances from their respective gas cells. Each star particle in FIRE represents a single-age, single-abundance stellar population, whose evolution is based on stellar population synthesis models from \textsc{STARBURST99} \citep{Leitherer_etal_1999} assuming a Kroupa initial mass function (IMF) \citep{Kroupa_2001}. FIRE also models stellar feedback processes like core-collapse and Typa Ia supernovae, radiation pressure, continuous mass loss, photoionization and photo-electric heating in a time-resolved manner. Furthermore, FIRE simulations implement a spatially uniform UV/X-ray ionizing background that evolves with redshift \citep{Faucher-Giguere_etal_2009}. Some of the caveats of this model are that it neglects the inhomogeneous nature of reionization and also has an earlier reionization time ($z \sim 10$) than recent measurements suggest ($z \sim 7$) \citep{Planck_2020}. But all the simulations that we use in this work self-consistently employ this model and hence will not result in any simulation-to-simulation difference due to these caveats. Additionally, these simulations includes spurious heating of neutral gas at $\lesssim$1000 K at $z \gtrsim 10$ due to cosmic rays, which suppresses star formation in low-mass halos at high redshifts. But this has no effect after reionization begins and hence our results at $z \lesssim 9$ are self-consistent \citep{Gandhi_etal_2024}. For a more detailed description of the physics implemented in FIRE simulations, see \citet{Hopkins_etal_2018a, Hopkins_etal_2018b}. 

The FIRE-2 simulations successfully generate isolated MW-like galaxies and MW/M31-like galaxy pairs, reproducing the observed properties of MW stellar disks, bars, and spiral structures \citep{Ma_etal_2017, Hopkins_etal_2018a, Sanderson_etal_2020, Ansar_etal_2025};
as well as stellar mass functions, radial distance distributions, and star formation histories of their satellite galaxies \citep{Wetzel_etal_2016, Garrison-Kimmel_etal_2019a, Garrison-Kimmel_etal_2019b, Samuel_etal_2020, Samuel_etal_2021}. The central massive MW-like galaxy that forms self-consistently in these simulations efficiently destroys satellite galaxies and low-mass sub-halos that orbit within the inner $\sim$30 kpc through gravitational tidal forces \citep{Garrison-Kimmel_etal_2017, Barry_etal_2023}. The resulting radial distribution of MW satellite galaxies is consistent with observations \citep{Samuel_etal_2020}. 
The tidal disruption of well-resolved simulated satellite galaxies also leads to a realistic population of stellar streams \citep{Panithanpaisal_etal_2021, arora2022stability,Horta_etal_2023,Shipp_etal_2023}, albeit with some discrepancies in their pericenter and apocenter distributions compared to observed MW low-mass galaxy streams \citep{Shipp_etal_2023}. However, streams from globular clusters do not naturally form in the FIRE-2 simulations due to their limited resolution. 

In this work, we make a catalog of surviving low-mass galaxies, stellar streams, and phase-mixed galaxies, along with their surrounding low-mass galaxies (field galaxies), around 13 MW-mass halos: seven isolated MW-like halos from the \textit{Latte} suite \citep{Wetzel_etal_2016}, and three LG-like MW/M31 pairs from the \textit{ELVIS on FIRE} suite \citep{Garrison-Kimmel_etal_2019a}. All simulations follow a $\Lambda$CDM cosmology, with cosmological parameters roughly consistent with the \citet{Planck_2020} results. Cosmological initial conditions are generated at $z$ = 99 using \textsc{MUSIC} \citep{Hahn_and_Abel_2011}; the zoomed regions are selected from periodic cosmological boxes of sizes 70--172 Mpc. The star/gas particles in the isolated Latte suite have an initial mass resolution of $\sim$7100 $M_\odot$, while in the paired simulations of the ELVIS suite, it is 3500--4000 $M_\odot$. The gravitational softening length (spatial resolution) for the star particles is fixed at $\sim 4$ pc, while the gas particles have an adaptive softening length with a minimum of $\sim 1$ pc. The dark matter (DM) particles have a resolution of 19000--39000 $M_\odot$ in mass, and a spatial resolution of $\sim$40 pc. The main host halos have a total mass of $M_{\rm 200m}$ $\approx$ 1--2 $\times$ 10$^{12}$ $M_\odot$\footnote{`200m' indicates a measurement relative to 200 times the mean matter density of the Universe} and a stellar mass of $M_\star$ $\approx$ 10$^{10-11}$ $M_\odot$, which spans measured values of MW's halo and stellar mass \citep{Sanderson_etal_2020}. 
In each simulation, 600 snapshots are saved from $z$ = 99 to $z$ = 0 with a typical spacing of $\sim$ 25 Myr after $z\sim 2$, allowing accurate tracking of disrupted galaxies.

Since an ultimate goal of the near-far technique is to infer the shape of the \emph{universal} SMF/UVLF at high redshifts, we also use results from the FIREbox$\mathrm{^{HR}}$ simulation \citep{Feldmann_etal_2024} as a benchmark. FIREbox$\mathrm{^{HR}}$, a higher-resolution re-run of FIREbox \citep{Feldmann_etal_2023}, is a cosmological volume simulation with a box size of 22.1 cMpc and with a mass resolution ($m_b \sim$ 7800 $M_\odot$) similar to the Latte suite. It contains $\sim$2000 galaxies brighter than $M_{\mathrm{UV}} = -14$ at $z \sim 6$, providing a good statistical sample. It uses the same FIRE-2 physics to model baryonic processes down to $z \sim$ 6, and encapsulates the multi-phase nature of ISM because of its high dynamic range and multi-channel stellar feedback. The simulated SMFs/UVLFs at $z \sim$ 6 - 14 are in good agreement with JWST observations \citep{Feldmann_etal_2024}. Furthermore, the observed cosmic UV luminosity density at these redshifts is well-reproduced by FIREbox$\mathrm{^{HR}}$. The simulation naturally explains recent JWST observations, which suggest a relatively high level of star formation activity even during early phases of the reionization era, supporting the accuracy of the baryonic processes implemented in FIRE-2. We use the SMFs/UVLFs from FIREbox$\mathrm{^{HR}}$ as our benchmark for the high-$z$ SMFs/UVLFs, and compare them against the ones derived from the progenitors of present-day substructures in the core suite. We also note that Gandhi et al. (in prep) compares their reconstructed SMFs/UVLFs to a universal SMF/UVLF from the statistical sample of galaxies from the FIRE-2 high-redshift zoom-in suite \citep{Ma_etal_2018a, Ma_etal_2018b, Ma_etal_2019} as a benchmark for the high-$z$ SMFs/UVLFs, which also been found to agree well with recent high-$z$ JWST observations \citep{Sun_etal_2023b}. Both sets of simulations show good agreement with each other with respect to the SMFs/UVLFs and cosmic UV luminosity density, though they may not match exactly.

\subsection{Identifying Low-mass Galaxies}

The DM halos and subhalos (smaller virialized halos within a larger halo) in each snapshot of the core simulations are identified using the \textsc{ROCKSTAR} 6D phase-space halo finder \citep{Behroozi_etal_2013a} to create a halo catalog for each of the 600 snapshots. The (sub)halos are identified according to their radius, such that it encloses 200 times the mean matter density of the Universe ($R_{\rm 200m}$), and only the ones with bound mass fraction $>$ 0.4 and at least 30 DM particles are kept. The bound mass fraction serves to select friends-of-friends groups in phase space that have sufficient density to be gravitationally self-bound. We take a value slightly lower than \textsc{ROCKSTAR's} default of 0.5 since we are interested in finding the nuclei of partially disrupted halos for further tracking, but well above the threshold of 0.15 where the halo mass function is sensitive to changes in this parameter \citep{Behroozi_etal_2013a}. 
After the halo catalogs have been formed, a merger tree is constructed by connecting the (sub)halos in time using \textsc{consistent-trees} \citep{Behroozi_etal_2013b}. To ensure numerical stability, the halo catalogs and merger trees are constructed using only DM particles \citep{Samuel_etal_2020}.

Star particles are then assigned to each (sub)halo in post-processing (adapted from the method described in \citealt{Necib_etal_2019} and \citealt{Samuel_etal_2020}), with slightly different criteria at $z = 0$ and $z \geq 6$:

\begin{enumerate}
    \item At $z = 0$, for each (sub)halo with radius $R_{\rm 200m}$ and center-of-mass velocity $V_{\rm circ, max}$ as obtained from \textsc{ROCKSTAR}, star particles are selected whose positions lie within $0.8 R_{\rm 200m}$ (out to a maximum radius of 30 kpc) and whose relative velocity falls within $2 V_{\rm circ, max}$. 

    \item Star particles whose positions are within $1.5\,R_{\rm 90}$ (the radius enclosing 90 percent of the mass of member star particles) 
    of \emph{both} the center-of-mass position of member stars and the DM halo center, and whose velocities are within twice the stellar velocity dispersion of the particles within the subhalo, are selected. 

    \item Steps (1) and (2) are repeated until the sum of the masses of all member star particles, $M_\star$, converges to within 1 percent. 

    \item Finally, all halos with at least six star particles and an average stellar density $>$ 300 $M_\odot$ kpc$^{-3}$ are selected.
    
\end{enumerate}

This method is conservative enough to keep galaxy properties stable across time at lower redshifts, and excludes nearly all stream stars \citep{Panithanpaisal_etal_2021, Samuel_etal_2020}. These assignments are then used as a starting point for identifying streams, as described in Section \ref{subsec:classification}. However, these criteria can be too stringent when assigning star particles to low-mass (sub)halos at $z \geq 6$, which in turn can lead to an incomplete low-mass (sub)halo population. In order to improve the overall completeness of the star particle assignment to each (sub)halo at $z \geq 6$, we follow the method implemented in \citet{Gandhi_etal_2024}: we change the enclosing radius from $0.8\,R_{\rm 200m}$ to $1.0\,R_{\rm 200m}$ and do not apply any constraints on the velocity relative to $V_{\rm circ, max}$. Furthermore, we keep all the halos with at least two star particles instead of six, since most of the galaxies that we analyze at high redshifts are progenitors of well-resolved galaxies at $z = 0$.

\begin{deluxetable*}{ccccccccc}
\tablenum{1}
\tablecaption{\textbf{Properties of the FIRE-2 Simulations at $z = 0$.} }
\label{tab:sim_data}
\tablewidth{0pt}
\tablehead{
\colhead{Simulation Name} & \colhead{$M_{\rm 200m}$} & \colhead{$M_{\rm \star, 90}$} & \colhead{$R_{\rm 200m}$} & \colhead{$z_{\mathrm{MR_{3:1}}}$} &
\multicolumn{4}{c}{Number of}\\
\cline{6-9}
\colhead{} & \colhead{($\times 10^{12} \, M_\odot$)} & \colhead{($ \times 10^{10} \, M_\odot$)} & \colhead{($\rm{kpc}$)} & \colhead{} & 
\colhead{\parbox[c]{1.5cm}{\centering \vspace{0.5em} Satellite \\ Galaxies}} & \colhead{\parbox[c]{1.5cm}{\centering \vspace{0.5em} Stellar \\ Streams}} & \colhead{\parbox[c]{2cm}{\centering \vspace{0.5em} Phase-mixed \\ Objects}} & \colhead{\parbox[c]{1.5cm}{\centering \vspace{0.5em} Field \\ Galaxies}} \\
}
\startdata
m12i\tablenotemark{A} & 1.2 & 6.3 & 336 & 3.18 & 20 & 13 & 17 & 8\\
m12b\tablenotemark{B} & 1.4 & 8.5 & 358 & 1.59 & 19 & 13 & 10 & 21\\
m12f\tablenotemark{C} & 1.7 & 7.9 & 380 & 3.77 & 28 & 12 &  29 & 23\\
m12c\tablenotemark{B} & 1.3 & 5.8 & 351 & 1.32 & 33 & 11 & 7 & 42\\
m12m\tablenotemark{D} & 1.6 & 11.0 & 371 & 1.68 & 31 & 13 & 6 & 32\\
m12r\tablenotemark{E} & 1.1 & 1.7 & 321 & 1.02 & 24 & 5 & 6 & 20\\
m12w\tablenotemark{E}& 1.1 & 5.7 & 319 & 2.92 & 31 & 5 & 14 & 24\\
\hline
Romeo\tablenotemark{B} & 1.3 & 6.6 & 341 & 5.70 & 23 & 20 & 37 & \multirow{2}{*}{74}\\
Juliet \tablenotemark{B} & 1.1 & 3.8 & 321 & 4.72 & 22 & 15 & 29 \\
\noalign{\vskip 5pt}
Romulus\tablenotemark{F}& 2.1 & 9.1 & 406 & 1.70 & 38 & 14 & 25 & \multirow{2}{*}{59}\\
Remus\tablenotemark{F}& 1.2 & 4.6 & 339 & 0.98 & 20 & 15 & 6 \\
\noalign{\vskip 5pt}
Thelma\tablenotemark{F} & 1.4 & 7.1 & 358 & 4.29 & 20 & 11 & 24 & \multirow{2}{*}{64}\\
Louise\tablenotemark{F}& 1.1 & 2.6 & 333 & 3.32 & 27 & 12 & 24 \\
\enddata
\tablecomments{\emph{Simulation Name}: References listed below; the top seven are isolated MW-like and bottom three are paired MW/M31-like. $M_{\rm 200m}$ and $R_{\rm 200m}$: total mass and spherical radius within which the mean density is 200 times the matter density of the Universe. $M_{\star, 90}$: stellar mass within a spherical radius that encloses 90\% of the stellar mass within 20 kpc. $z_{\mathrm{MR_{3:1}}}$: the redshift at which the stellar-mass ratio of the most massive halo to the second most massive subhalo is 3:1, which is when the dominant host galaxy emerges. Numbers of surviving satellite galaxies ($M_\star \gtrsim 10^{4.5} M_\odot$), stellar streams ($M_\star \gtrsim 10^{5.5} M_\odot$), phase-mixed galaxies ($M_\star \gtrsim 10^{5.5} M_\odot$), and field galaxies ($M_\star \gtrsim 10^{4.5} M_\odot$) are determined as described in Section \ref{subsec:classification}.}
\tablerefs{(A) \citet{Wetzel_etal_2016}, (B) \citet{Garrison-Kimmel_etal_2019a}, (C) \citet{Garrison-Kimmel_etal_2017}, (D) \citet{Hopkins_etal_2018a}, (E) \citet{Samuel_etal_2020}, (F) \citet{Garrison-Kimmel_etal_2019b}.}
\end{deluxetable*}

\section{Present-day Substructures and their High-redshift Progenitors} \label{sec:substuctures_and_progs}

\subsection{Substructure Classification Algorithm} \label{subsec:classification}

We classify present-day substructures into four categories:

\begin{enumerate}
    \item \textbf{Surviving satellite galaxies:} Intact low-mass galaxies within the virial radius (radii) 
    of the isolated (paired) MW-like (MW/M31-like) host(s) at $z = 0$. In this work, we consider the virial radius to be $R_{\rm 200m}$.

    \item \textbf{Stellar streams:} Disrupted low-mass galaxies which are torn and stretched along their orbits by tidal forces from the host galaxy. 

    \item \textbf{Phase-mixed galaxies:} Disrupted low-mass galaxies which were once stellar streams, but are now merged with the host galaxy and share its equilibrium phase-space distribution.

    \item \textbf{Field galaxies:} Intact galaxies that are outside the virial radius (radii) of the isolated (paired) MW-like (MW/M31-like) host(s) at $z = 0$, but within a maximum distance as discussed later.
\end{enumerate}

For the intact galaxies (satellite + field) at $z$ = 0, we select galaxies with stellar mass $M_\star \geq$ 10$^{4.5}$ $M_\odot$ (or number of star particles $\gtrsim$ 6) within 2 Mpc of the center of each isolated MW-like galaxy, or from the geometric center of mass of the LG-like pairs. These criteria ensure that our sample does not suffer from numerical issues due to very low-mass subhalos, and also avoid contamination by low-resolution DM particles \citep[same as in][]{Gandhi_etal_2024}. We also exclude star particles associated with the $z=0$ host galaxies themselves.

\begin{figure*}[ht!]
\centering
\includegraphics[width=\textwidth]{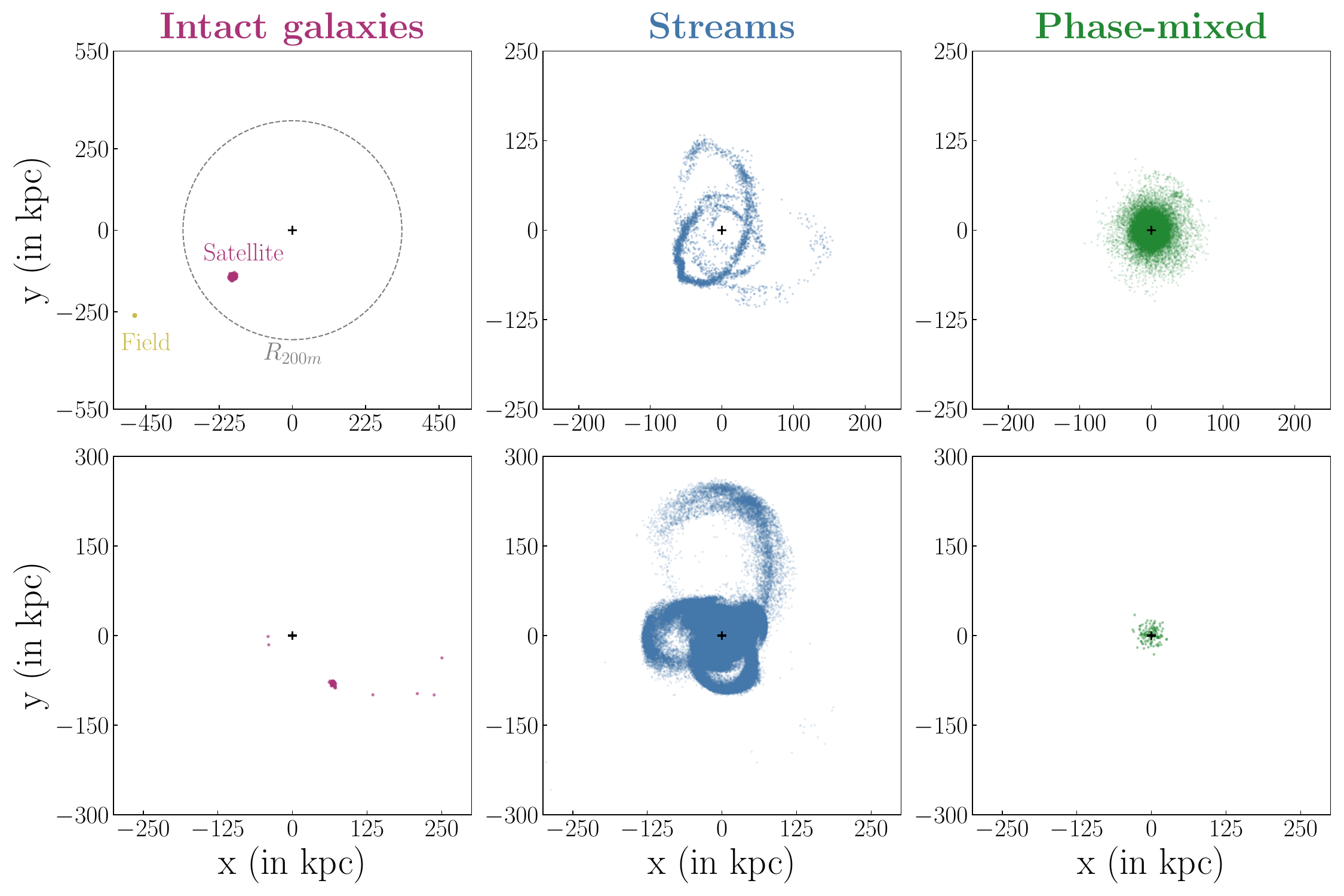}
\caption{\textbf{Present-day substructures classified by our algorithm, with manual reclassification.} The center of the host galaxy is denoted by the black cross in each panel. \textit{\textbf{Top:}} Three examples correctly classified by the automated algorithm described in Section \ref{subsec:classification}. \textit{ Left}: a satellite within $R_{\rm 200m}$ (black dashed circle) and a field galaxy outside $R_{\rm 200m}$ \textit{Center}: a stellar stream. \textit{Right}: a phase-mixed galaxy. \textit{\textbf{Bottom:}} Manual reclassification of substructures with complex morphologies. \textit{Left}: a satellite galaxy which was classified as a stream by the algorithm due to a handful of unbound star particles. \textit{Center}: a stream which was classified as a phase-mixed galaxy due to its large mass. \textit{Right}: a phase-mixed galaxy which was classified as a satellite galaxy due to its small size.} 
\label{fig:1}
\end{figure*}

We use a modified version of the method described in \citet{Panithanpaisal_etal_2021} to identify the satellites, streams, and phase-mixed galaxies in each of the halos. We track each bound subhalo with non-zero stellar mass within the virial radius of the main host at $z$ = 0 ($\sim$350 kpc). We start tracking the subhalos from the redshift ($z_{\mathrm{MR_{3:1}}}$) values calculated in \citet{Horta_etal_2024}. This is the redshift at which the stellar mass ratio of the most massive halo to the second most massive subhalo is 3:1 (see also \citealt{Santistevan_etal_2020}), and is used to define the time when the dominant host galaxy emerges. The subhalos that merge with the main progenitor galaxy before this redshift are considered to be part of the host galaxy and we do not include them. This is the earliest point in time that we can reasonably distinguish the host from satellites. Table \ref{tab:sim_data} lists $z_{\mathrm{MR_{3:1}}}$ for each of the host halos. We also update the selection of member stars from \citet{Panithanpaisal_etal_2021}. Following \citet{Shipp_etal_2023}, we require that each stream member star belongs to its progenitors for at least 10 snapshots. This removes contamination from unassociated satellites and from host disk stars that may be erroneously picked up by the halo finder due to coincidental proximity in phase-space.

We use three criteria to select and classify present-day substructures. We first require that the number of star particles be larger than 100 ($M_*\gtrsim 5 \times 10^5\,M_\odot$).\footnote{We manually reduce the stellar mass requirement to $10^{4.5}$ $M_\odot$ for the case of satellite galaxies to be consistent with \citet{Gandhi_etal_2024}. This choice seems reasonable since these galaxies are more compact than streams and therefore easier to resolve and detect. In Section \ref{sec:detectability} we quantify detectability in terms of surface brightness.} This ensures that the streams have enough star particles to be resolved in our simulations. To be considered disrupted rather than intact, the maximum pairwise separation between any two star particles must also be greater than 120 kpc. Finally, the local velocity dispersion $\sigma$ of each disrupted substructure is compared to a stellar-mass dependent threshold $\sigma_{\mathrm{th}}(M_*)$ \citep[][Eq. 2]{Panithanpaisal_etal_2021}. Substructures with $\sigma<\sigma_{\mathrm{th}}(M_*)$ are considered streams, while those with  $\sigma>\sigma_{\mathrm{th}}(M_*)$ are considered phase-mixed galaxies. In some borderline cases where a substructure barely satisfies some criterion, but looks distinctively like another category, we reclassify the substructure manually. Table \ref{tab:sim_data} shows the number of different substructures in each of the FIRE-2 simulations, along with the properties of the host(s). We show examples of the substructures classified by the algorithm as satellite galaxies, stellar streams, phase-mixed galaxies, and field galaxies, in Figure \ref{fig:1}, along with cases where we manually reclassify the original assignment according to the algorithm.

Unlike \citet{Panithanpaisal_etal_2021}, we do not set any upper bound on the number of star particles, which leads us to include very massive objects ($M_\star$ $\sim$ 5 $\times$ 10$^9$ $M_\odot$). Clearly, the progenitors of these substructures also contribute to the UV and ionizing photon budget during reionization. However, these disrupted systems can exhibit complex morphologies, with some parts displaying a coherent tidal tail and other parts appearing more phase-mixed (Figure \ref{fig:1}, bottom center panel). Most of these cases are classified as phase-mixed according to our criteria, but if the object has a prominent tidal tail, we manually reclassify it as a stream, as was done for the stream in Figure \ref{fig:1}. 

Another common scenario for manual reclassification is shown in the bottom left panel of Figure \ref{fig:1}. This satellite galaxy was originally classified as a stream by the algorithm because of a single pair of particles separated by more than 120 kpc in distance. We reclassify it as a low-mass satellite galaxy since the bound part clearly dominates; such a substructure would almost certainly be detected as a satellite. 

Finally, in some low mass phase-mixed galaxies, the local velocity dispersion is below the threshold, while the pairwise particle separations are smaller than 120 kpc. These objects are classified as satellites by the algorithm, but their star particles are distributed around the center of the host galaxy (Figure \ref{fig:1}, bottom right panel). These are manually reclassified as phase-mixed. 

We stress that our classifications in this work are based on the entire distribution of simulated star particles. A more robust classification requires taking observability effects into the picture as in \citet{Shipp_etal_2023}, which can partially mitigate the problem of these morphologically complex substructures. We will explore this in future work.

\subsection{Identifying the High-Redshift Progenitors} \label{subsec:track_high-z}

We track the $z = 0$ star particles assigned to present-day substructures back to $z \gtrsim 6$ and determine their progenitor host halos from the corresponding halo catalogs. We assign a galaxy (a halo containing star particles) as a progenitor of a substructure if at least one of its star particles ends up in that substructure at the present day. We do not include the progenitors of the present-day central MW/M31-mass galaxies. We consider a progenitor galaxy to be a high-redshift contributor to the proto-MW/LG if it has at least one star particle that ends up within 2 Mpc from the center of the host galaxy (or the geometric center of mass of the paired hosts) at $z = 0$. 

Star particles formed in the same progenitor galaxy at high redshift can end up in different galaxies at the present day. For example, a large fraction of the star particles from a progenitor galaxy may end up in a particular stream,
while a smaller fraction may end up in a second stream, usually due to a close dynamical interaction during its evolution. In such cases, if a progenitor contributes stars to two $z=0$ objects with the same classification, it is counted once.
If a single progenitor contributes to multiple objects with different present day classifications (e.g., a stream and a phase-mixed galaxy), then the progenitor is counted as a predecessor of each object. However, when we count progenitors among multiple different $z=0$ categories (e.g., stream {\em and} phase-mixed galaxies), we count each progenitor only once to avoid double-counting.

A small number of star particles in the $z = 0$ substructures exist at $z \gtrsim 6$ but are not associated with any progenitor galaxy in the halo catalogs. Since this is a small fraction of the total stellar mass, we ignore these star particles when we assign progenitors. There are also a small number of $z=0$ galaxies containing star particles that can be traced back to $z \gtrsim 6$ but are not assigned to any galaxy at this epoch. These are mainly the lowest-mass galaxies close to the limit of our resolution. Like \citet{Gandhi_etal_2024}, we assert that each of these galaxies has one progenitor at that redshift, and assign all the traceable stellar mass to this single progenitor.

\begin{figure*}[ht!]
\centering
\includegraphics[width=\textwidth]{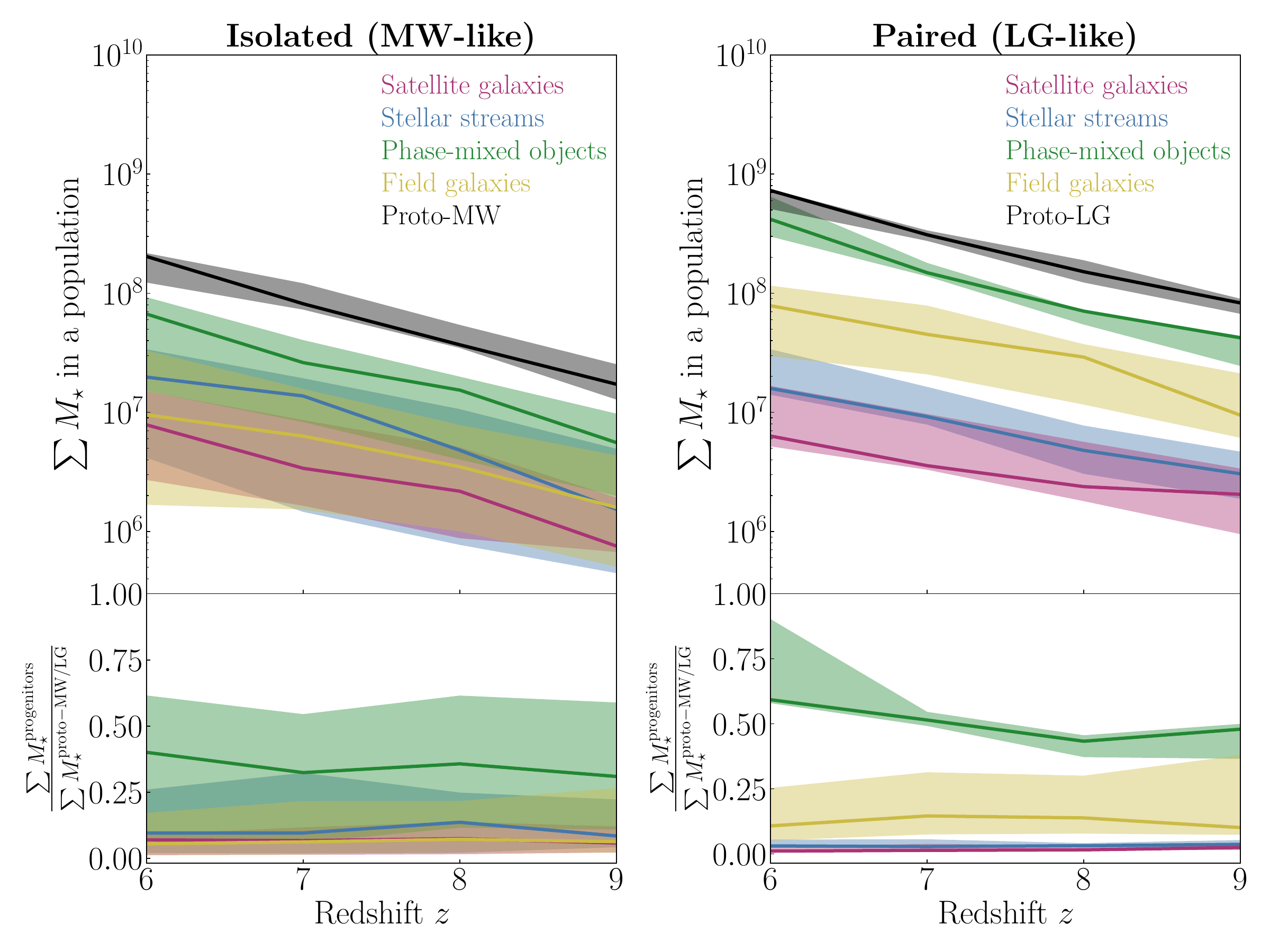}
\caption{\textbf{Streams and phase-mixed galaxies contribute more to the stellar mass budget at the EoR than intact low-mass galaxies.} In all panels, solid lines show the median across the simulations and shaded regions show the 68\% halo-to-halo scatter, for the isolated (left) and the paired (right) galaxy suites. The red, blue, green, and yellow lines show the total stellar mass of the progenitors of satellite galaxies, stellar streams, phase-mixed galaxies, and field galaxies, respectively. The black line shows the stellar mass of the proto-MW/LG, defined here as the total stellar mass budget from all progenitors, including those associated to the present-day main host galaxy (or host galaxies, in the paired case). \textit{\textbf{Top}}: total stellar mass at EoR for each type of substructure.  \textit{\textbf{Bottom}}: fractional contribution to the stellar mass from each substructure type (i.e., the ratio of the colored lines in the top panel to the solid black line).} 
\label{fig:2}
\end{figure*}

\begin{figure*}[ht!]
\centering
\includegraphics[width=\textwidth]{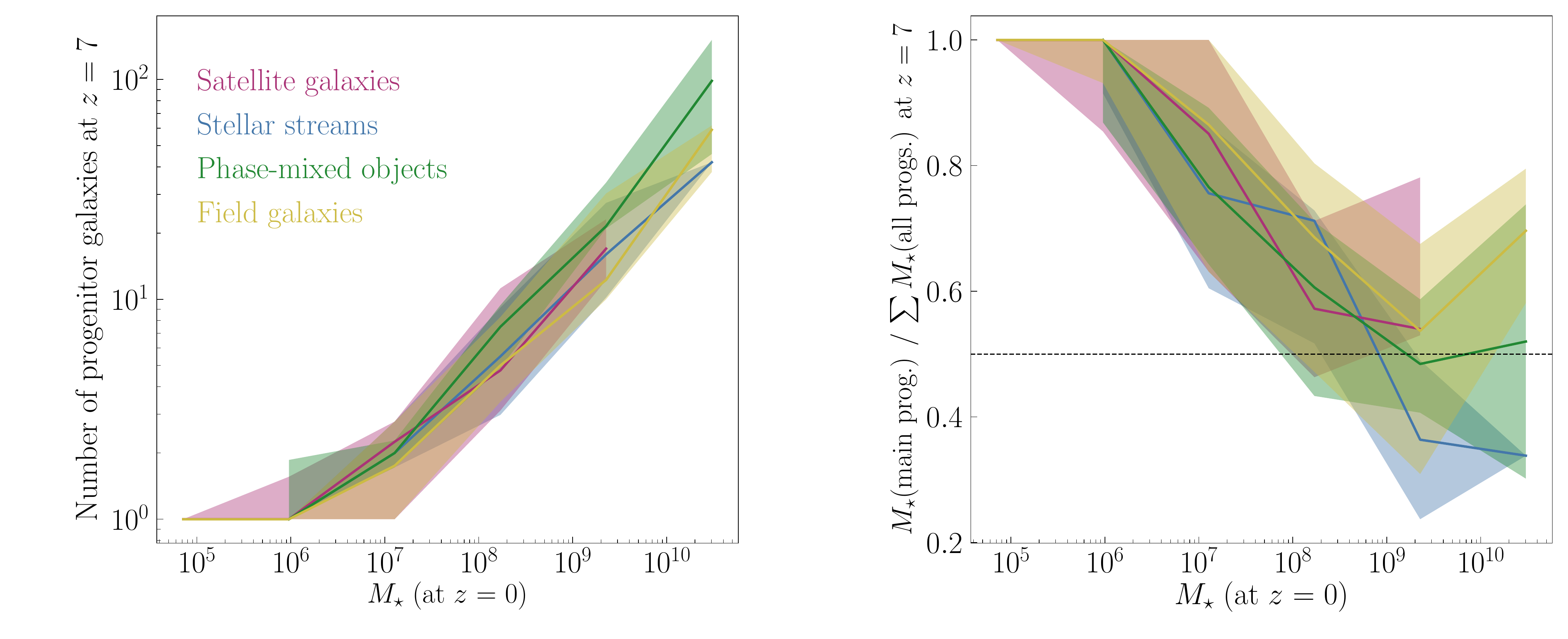}
\caption{\textbf{Hierarchical assembly at high redshifts is similar for all substructure types.} In both panels, solid lines show the median across all simulations (both isolated and paired) at $z=7$; shaded regions show 68\% halo-to-halo scatter. \textit{\textbf{Left}}: number of progenitors of satellite galaxies (red), stellar streams (blue), phase-mixed galaxies (green), and field galaxies (yellow) as a function of present-day stellar mass. The number of progenitors at high redshifts increases with $z=0$ stellar mass. \textit{\textbf{Right}}: fractional contribution of the main progenitor to total progenitor stellar mass. The main progenitor dominates the fractional contribution for masses between $M_\star$ = 10$^{4.5-9}$ $M_\odot$, with $\sim$100\% contributions at $M_\star$ $\sim$ 10$^{5-6}$ $M_\odot$.}
\label{fig:3}
\end{figure*}

\section{Stellar mass distribution of progenitor galaxies at high redshifts} \label{sec:mass_budget_and_struct_form}

Having assigned progenitor galaxies to present-day substructures, we next analyze the contributions of each classification to the total stellar mass distribution of the proto-MW/LG during the reionization epoch. We then reconstruct the high-$z$ SMFs/UVLFs and determine the effect on their amplitude and shape of including the present-day disrupted galaxies. 

\subsection{Stellar-mass Budget During the Epoch of Reionization}
\label{subsec:stellar_mass_budget}

The amount of stellar mass contained by the proto-MW/LG and formed before $z\sim 6$, often referred to as the ``stellar mass budget'' at the EoR, controls the number of photons available to reionize the Universe. Thus, we first explore how the total proto-MW/LG stellar mass is distributed among the progenitors of each present-day classification. 

Figure \ref{fig:2} shows the total progenitor stellar mass budget at $z=6-9$ from each classification, as well as their fractional contributions to the stellar mass of the proto-MW/LG. We show the results from the isolated (MW-like) and paired (LG-like) galaxy suites in the left and right panels, respectively, since the number of galaxies is different in each suite. Each line shows the median stellar mass budget across all of the simulations in that particular suite, while the shaded region shows the 68\% halo-to-halo scatter. The black line indicates the total stellar mass from all of the progenitors (proto-MW/LG); the red, blue, green, and yellow lines show the contributions from the progenitors of present-day satellite galaxies, stellar streams, phase-mixed galaxies, and field galaxies, respectively. We define the proto-MW by including the progenitors of the present-day host galaxy in addition to the classified substructures (up to 2 Mpc from the center of the host galaxy). For the proto-LG, we include the progenitors of both hosts along with the classified substructures (up to 2 Mpc from the geometric center of mass of the two hosts).  

In the isolated suite, the progenitors of phase-mixed galaxies contribute the most to the total stellar mass budget, followed by the stellar stream progenitors. In the paired suite, progenitors of phase-mixed galaxies again contribute the most to the total stellar mass budget, followed by the progenitors of field galaxies, streams, and satellites, respectively. Though the stream progenitors contribute less than that of the field galaxies in the paired suite, in both cases, the progenitors of the present-day disrupted galaxies yield more stellar mass during the EoR than the progenitors of intact low-mass galaxies. Even though the stream progenitors contribute less than phase-mixed galaxies, they play a larger role than intact low-mass galaxies. The fractional stellar mass contribution from progenitors of each classification, relative to progenitors of the main galaxy, is roughly constant from $z=6-9$, with disrupted galaxies contributing at least as much as intact satellites across this entire range (Figure \ref{fig:2}, bottom panels). There is no strong mass-dependent bias across different present-day substructures. This suggests that neglecting the progenitors of streams and phase-mixed galaxies can underestimate the reionization-era ionizing photon budget derived from the near-far technique. However, even with adding the disrupted galaxies we can recover only up to $\sim$50\% of the total budget. The rest of the stellar mass is encompassed by the progenitors of the present-day main host galaxy \citep{Horta_etal_2024}. 

The field galaxy progenitors contribute more to the stellar mass budget than the stream progenitors in the paired suite because of our present-day distance selection criterion (2 Mpc). In the paired simulations, the cumulative number of low-mass galaxies beyond 1 Mpc continues to increase faster than in the isolated suite (Gandhi et al., in prep), leading to more field galaxies (Table \ref{tab:sim_data}) and hence a greater contribution from field galaxy progenitors at high redshift. If we instead use a distance cutoff of 1 Mpc, we find consistent behavior between the isolated and paired galaxy suites.

\subsection{Hierarchical Structure Formation at High Redshifts}
\label{subsec:galaxy_assembly}

The near-far technique assumes that the stellar mass inferred from a present-day fossil record came from a single progenitor at all redshifts. In \citet{Gandhi_etal_2024} and in this work, we similarly define the {\em fossil record} by combining the total stellar mass from all the progenitors of each present-day structure into one. We also consider instead assigning each present-day galaxy to its main progenitor, defined to be the progenitor with the highest stellar mass. We discuss the main progenitor only in this section, for purposes of illustration, and proceed with the fossil record assumption in the rest of the paper when assuming a single progenitor. 

The viability of the fossil record assumption is important because the reconstructed SFHs alone do not directly supply the complete merger histories of present-day galaxies. However, structure formation proceeds hierarchically (at least in cold dark matter models): present-day substructures were generally built up through mergers, so their stellar mass at high redshift may be distributed across multiple progenitors. Thus, the fossil record assumption could potentially bias inferences of the high-$z$ SMFs/UVLFs.

\citet{Gandhi_etal_2024} explored the effects of mergers and disruptions for present-day surviving low-mass galaxies by quantifying their progenitor counts during the reionization era. 
They found that the lowest mass simulated galaxies ($M_\star$ $\sim$ $10^5$ $M_\odot$) have $\sim$1--2 progenitors at $z\gtrsim6$, while the progenitor number increases with present-day stellar mass, with the most massive galaxies ($M_\star$ $\sim$ $10^9$ $M_\odot$) having $\sim$15--40 progenitors. Furthermore, they showed that the fractional contribution of the main progenitor galaxy relative to the total stellar mass from all progenitor galaxies at that redshift, is $\gtrsim 50\%$ dominant across a wide range of present-day stellar masses ($M_\star$ $\sim$ 10$^{4.5-8}$ $M_\odot$). The present-day lowest-mass galaxies have an order unity contribution from a single main progenitor since they have at most $\sim$1--2 progenitors. On the other hand, high-mass present-day substructures are typically assembled from many smaller units, and the main progenitor makes a smaller fractional contribution to the total progenitor stellar mass at high redshift. Nevertheless, the main progenitor of even the most massive galaxies contributes $\sim$50\% to the total progenitor stellar mass. Ultimately, \citet{Gandhi_etal_2024} concluded that the assumption of a single progenitor in case of the surviving low-mass galaxy fossil record still allows an accurate recovery of the slopes at the low-mass/faint end of the high-$z$ proto-MW/LG SMFs/UVLFs.

Stellar streams and phase-mixed galaxies have different dynamical and formation histories than intact low-mass galaxies \citep{Panithanpaisal_etal_2021,Cunningham_etal_2022}. Thus, it is not clear that the fossil record assumption will extend trivially to these other classifications even if it works for intact galaxies. To confirm that disrupted galaxies also follow the fossil record, we examine the hierarchical formation histories of the different types of present-day substructures back to $z = 7$ (Figure \ref{fig:3}). We consider both the number of progenitors at $z = 7$ as a function of their present-day stellar mass (left panel), and the fractional contribution of the main progenitor (the ratio of the main progenitor's stellar mass to the total mass from all progenitors evaluated at $z = 7$; right panel), as a function of present-day stellar mass. Since these relations are similar for the isolated and paired simulations, we show the median trend over all 13 halos, with the shaded regions denoting the 68\% halo-to-halo scatter. 

The progenitors of stellar streams (blue) and phase-mixed galaxies (green) follow a pattern of hierarchical assembly that is indistinguishable from that of satellite (red) and field galaxies (yellow). As for intact galaxies, the least massive streams and phase-mixed galaxies ($M_\star \sim 10^{6-7} \, M_\odot$ at the present day) have $\sim$1--2 progenitors at $z = 7$; while the most massive ones ($M_\star \sim 10^{9-10} \, M_\odot$) have $\sim$30--70 progenitors. The slope of these curves is $\approx$0.38 for all classifications. The main progenitors of the present-day streams and phase-mixed galaxies (right panel) are also the greatest contributors to the stellar mass budget at $z = 7$ across a wide range of present-day stellar masses ($M_\star$ = 10$^{4.5-9}$). Although not shown here, we find similar behavior at $z$ = 6, 8, and 9. Thus, the assumption of a single main progenitor appears as accurate for the streams and phase-mixed galaxies as it is for the intact low-mass galaxies.

\section{Stellar Mass Functions and Ultraviolet Luminosity Functions at High-Redshifts} \label{sec:SMFs_and_UVLFs}

We further test the fossil-record assumption by reconstructing the high-$z$ SMFs/UVLFs including disrupted galaxies. We examine how well the SMFs/UVLFs of the proto-MW/LG (and by extension the universal SMF/UVLF) at $z \gtrsim 6$ can be recovered from each present-day classification category, as well as from several different combinations of intact and disrupted galaxies. For now, our analysis assumes that the SFHs of these substructures can be precisely reconstructed from their CMDs at $z = 0$. We compute the reconstructed SMFs/UVLFs in both the isolated and paired suite, and compare with the SMF/UVLF from FIREbox$\mathrm{^{HR}}$, which we treat as a benchmark for the universal SMF/UVLF. We do this for both the differential and cumulative functions.

We determine the SMFs of high-$z$ progenitor galaxies by counting the number of halos in logarithmic mass bins. In order to account for the increasing number of halos at small masses and to reduce the noise at the high-mass end, we adopt variable bin widths: $0.25$ dex over $[4.5, 5.5]$, $0.5$ dex over $[5.5, 6.75]$, and $0.75$ dex over $[6.75, 9.0]$. This adaptive binning scheme ensures adequate sampling across the full dynamic range in mass: these bin widths capture the shape of the SMFs, without excessive noise. We then divide by a volume factor, which we calculate by constructing a convex hull using all the halos at that redshift, including non-luminous ones, to avoid selection bias and account for voids to the extent permitted by simulation volume. This volume is distinct from 
the periodic box size used to calculate the FIREbox$\mathrm{^{HR}}$ SMFs/UVLFs. Since FIREbox$\mathrm{^{HR}}$ is a non-zoomed cosmological simulation with periodic boundary conditions, it has a well-defined total volume; for zoom-in simulations, which do not have periodic boundary conditions, the convex hull is the closest equivalent \citep{DREAMS_2025}.

The main goal of the near-far technique is to probe the faint-end of the UVLF during reionization. Hence, in addition to the SMFs, we also assess the impact of the progenitors of present-day disrupted galaxies on the reconstructed high-$z$ UVLF. Following the approach in \citet{Sun_etal_2023a}, we compute the rest-frame UV magnitude ($M_{\mathrm{UV}}$) at 1500 $\text{\r{A}}$ for each progenitor galaxy at high redshift. We then construct the differential and cumulative UVLFs for the same set of progenitor populations as the SMFs. We follow the same technique used to construct the SMFs, but now we count the number of progenitor galaxies in bins of UV magnitudes. Again, we adopt variable bin widths: $-0.75$ dex over $[-8.00, -10.25]$, $-1.50$ dex over $[-10.25, -14.75]$, and $-2.00$ dex over $[-14.75, -21.75]$.

These procedures result in reconstructed high-$z$ SMFs/UVLFs for both the \textit{all-progenitors} and \emph{fossil-record} scenarios. We compare them to universal benchmark SMFs/UVLFs in sections \ref{subsec:SMFs_and_UVLFs_all_progs} and \ref{subsec:SMFs_and_UVLFs_fossils}.

\begin{figure*}[ht!]
\centering
\includegraphics[width=\textwidth]{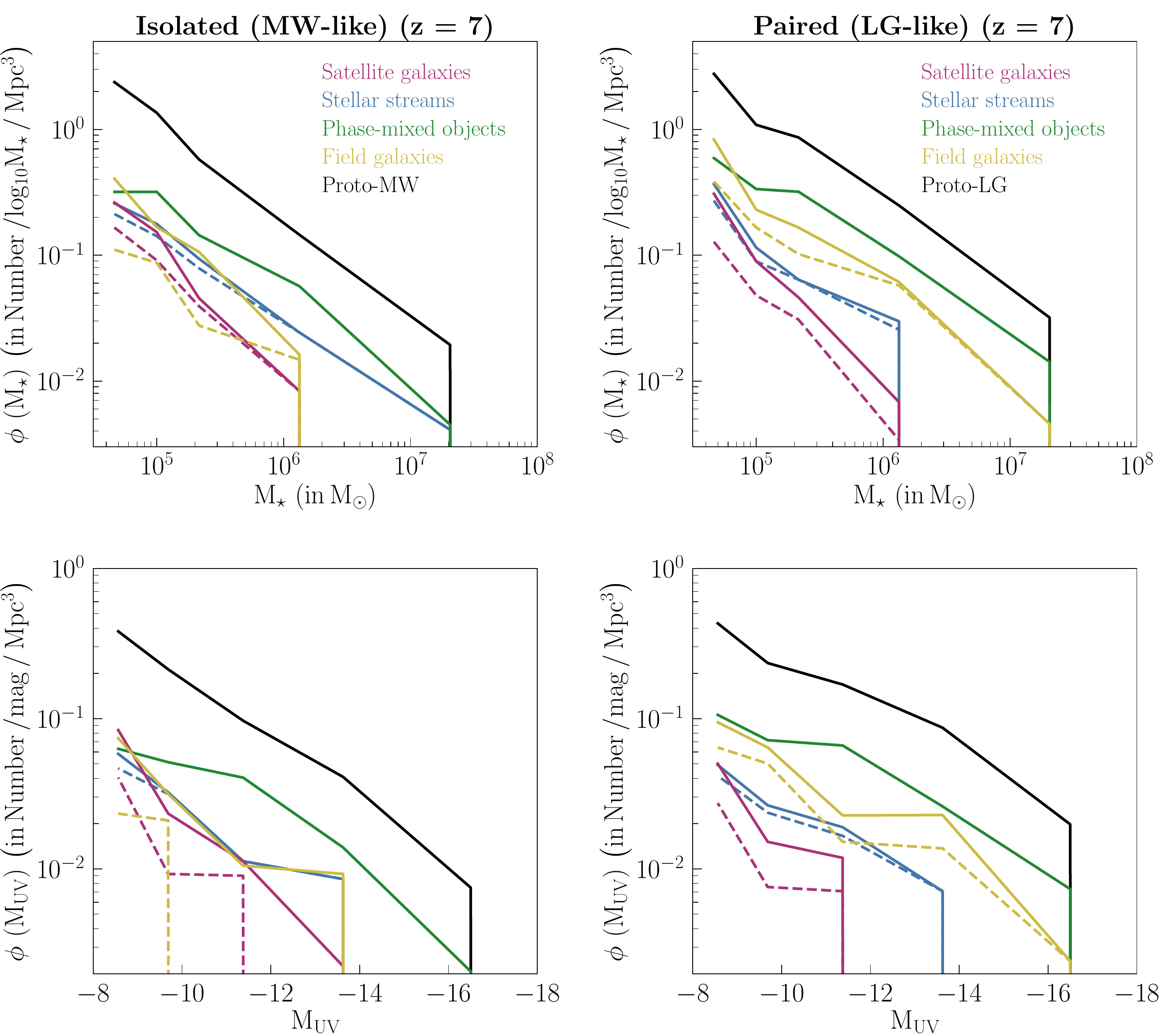}
\caption{\textbf{Present-day disrupted and intact galaxies recover different differential SMFs and UVLFs at $z=7$}. 
All panels show median reconstructed SMFs (top) and UVLFs (bottom) for the progenitors of various present-day substructure types, without (solid curves) and with (dashed curves) detectability constraints (Section \ref{sec:detectability}). The isolated galaxy suite is shown in the left column; paired galaxies in the right column. The proto-MW/LG (black), which includes the progenitors of the present-day main host(s), is compared with reconstructions from the progenitors of satellite galaxies (red), stellar streams (blue), phase-mixed galaxies (green), and field galaxies (yellow). 
} 
\label{fig:4}
\end{figure*}

\subsection{Reconstructed Differential Stellar Mass Functions and Ultraviolet Luminosity Functions for Each Present-day Classification} \label{subsec:SMFs_and_UVLFs_each_class}

Figure \ref{fig:4} compares the differential SMFs (top panel) and UVLFs (bottom panel) of the proto-MW/LG (black) at $z = 7$ with the ones constructed from the progenitors of present-day satellite galaxies (red), stellar streams (blue), phase-mixed galaxies (green), and field galaxies (yellow). The solid lines show reconstructions from the full set of progenitors and the dashed lines show reconstructions obtained using a simple model for the detectability of present-day substructures with the LSST (section \ref{sec:detectability}). 
We show results for the isolated suite and the paired suite in the left and right panels, respectively. The comparison between the two panels illustrates how well the reconstructed SMFs/UVLFs from each classification compare with the proto-MW/LG SMF/UVLF, when the near-far technique is applied to each of the MW and the LG alone. 
The curves show the median across all of the simulations in the respective galaxy suite. To highlight differences between classifications, the SMFs/UVLFs are computed in the \textit{all-progenitors} scenario rather than the \emph{fossil-record} scenario. 

The shape of the SMFs/UVLFs in the low-mass/faint end ($M_\star \lesssim 10^6 M_\odot$ or $M_{UV} \gtrsim -13$) reconstructed from the progenitors of present-day stellar streams and phase-mixed galaxies is qualitatively similar 
in representing in the proto-MW/LG SMFs/UVLFs, when compared to the SMFs/UVLFs from just the progenitors of intact galaxies. Furthermore, the disrupted galaxy progenitors recovers more of the normalization of the SMFs/UVLFs at least in comparison to those from intact present-day satellite galaxies. This underlines the need to include present-day disrupted galaxies in near-far analysis. The SMFs/UVLFs from the disrupted galaxies tend to extend to more massive/brighter galaxies, because more of these systems have massive progenitors at high redshifts than in the intact galaxy population. 
The larger normalization of the SMFs/UVLFs from these disrupted systems, compared to that recovered from the low-mass satellites alone, is partly because some of these massive systems have multiple high-$z$ progenitors.
The overall shapes do not change drastically even when detectability constraints are imposed, suggesting that it may be feasible to include disrupted galaxies in future near-far technique analyses.  
The SMFs and UVLFs reconstructed from the stream progenitors remain almost unchanged, because the detectable streams are the most massive ones with the largest number progenitors, comprising a large portion of the distribution at high redshifts.

\subsection{Reconstructed Differential Stellar Mass Functions and Ultraviolet Luminosity Functions Using the Full Merger Histories of Present-day Substructures} \label{subsec:SMFs_and_UVLFs_all_progs}

\begin{figure*}[ht!]
\centering
\includegraphics[width=\textwidth]{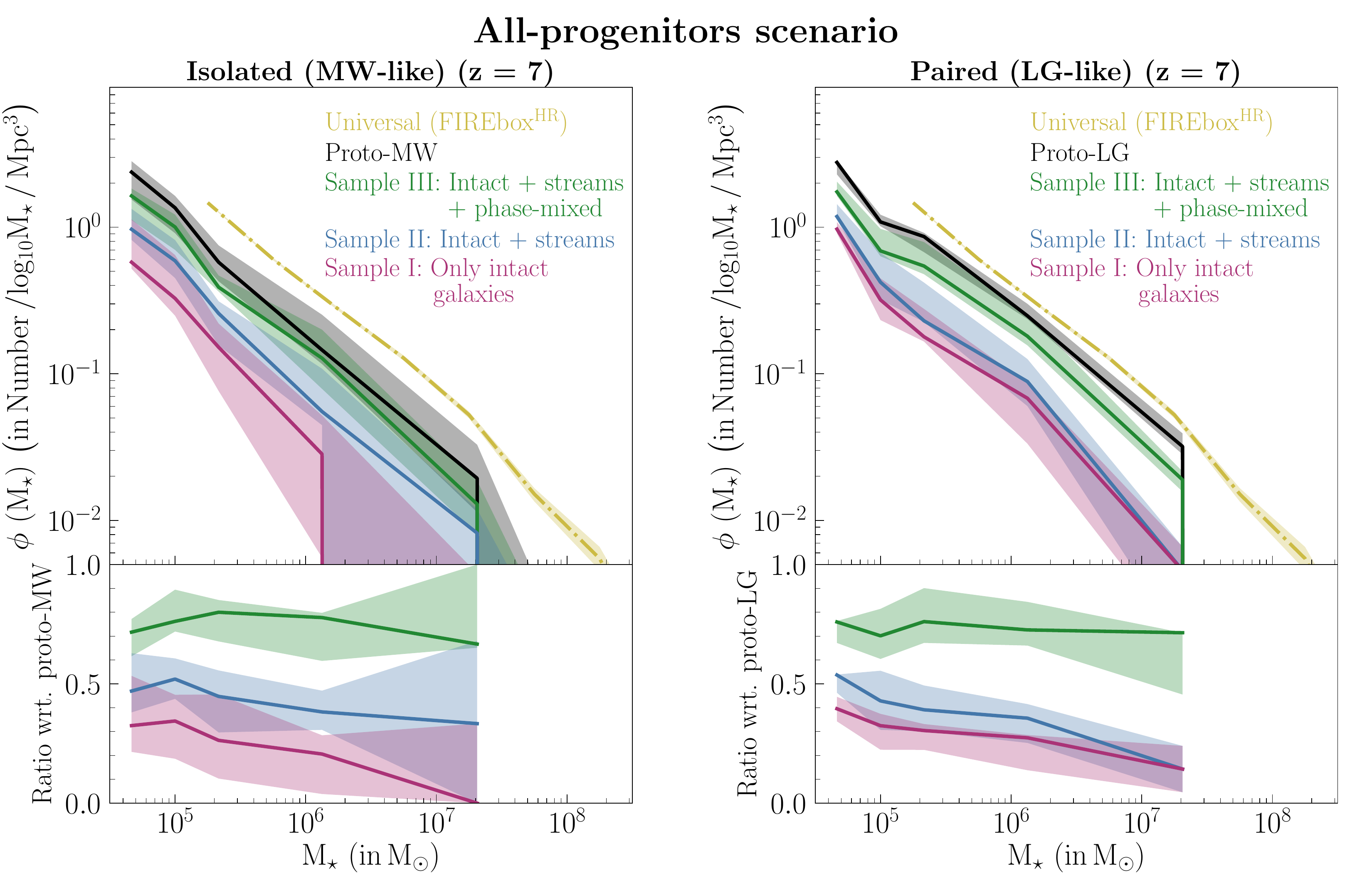}
\caption{\textbf{Differential SMFs at $z = 7$ in the \emph{all-progenitors} scenario.} In all the panels, red, blue, and green solid lines show the median across all the simulations for Samples I, II, and III, respectively in the isolated (left) and paired (right) galaxy suites. The shaded regions show the corresponding 68\% halo-to-halo scatter. In this scenario we measure all progenitors separately, instead of treating all progenitors as a single galaxy, as in the case of the fossil record. \textbf{\textit{Top}}: reconstructed differential SMFs at $z = 7$ from all of the progenitors in each present-day sample, compared to the proto-MW/LG (black; includes the present-day main host(s)) and the universal differential SMF from FIREbox$\mathrm{^{HR}}$ (dashed yellow). 
\textbf{\textit{Bottom}}: ratio of the SMF from each sample to the SMF of the proto-MW/LG. Reconstructed SMFs from sample II and III represent both the shape and normalization of the proto-MW/LG SMF better than sample I at all stellar masses, and generally appears to have smaller halo-to-halo scatter with some exceptions (see text for details).
} 
\label{fig:5}
\end{figure*}

\begin{figure*}[ht!]
\centering
\includegraphics[width=\textwidth]{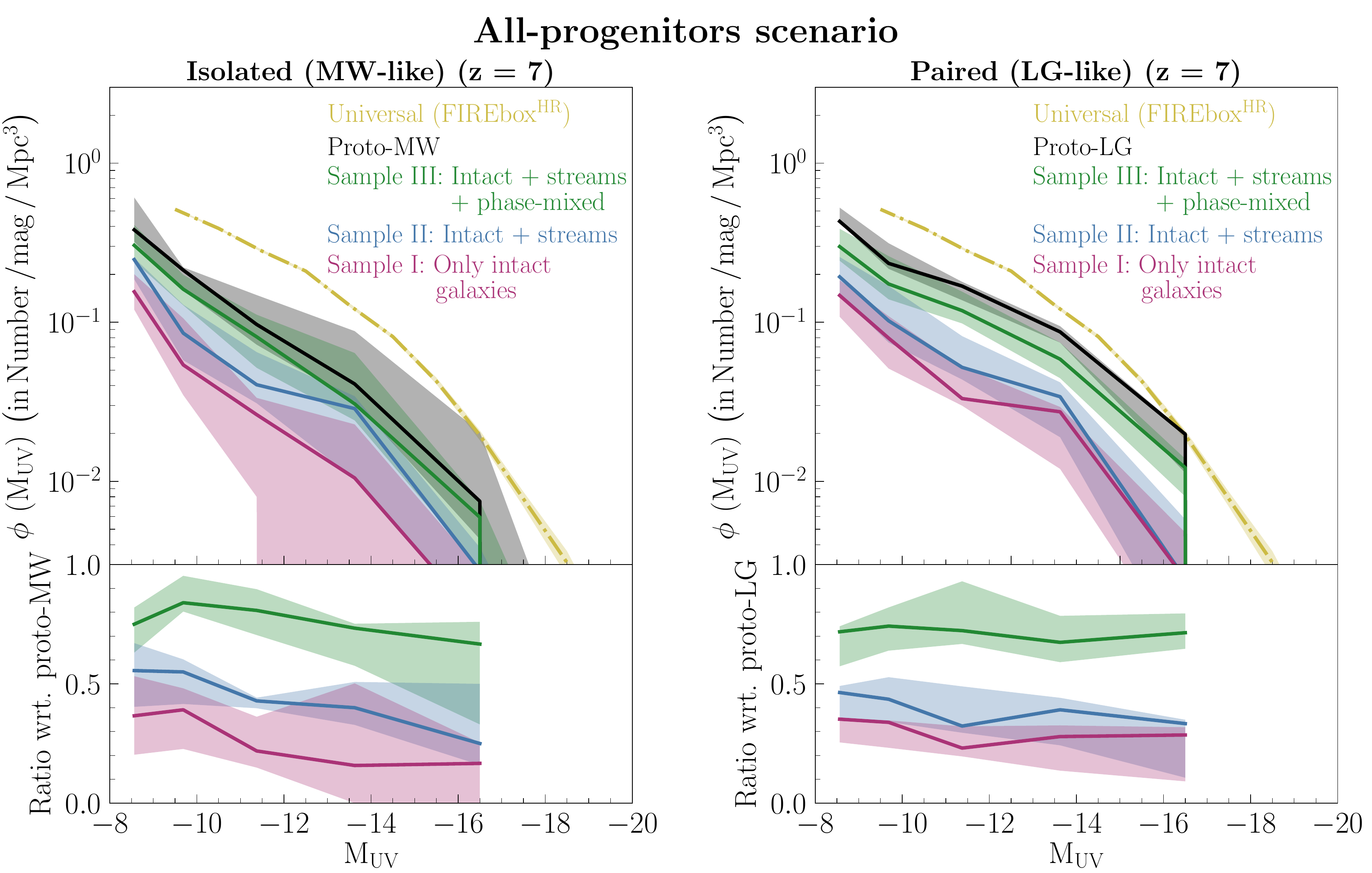}
\caption{\textbf{Differential UVLFs at $z = 7$ in the \emph{all-progenitors} scenario.} Same as Figure \ref{fig:5}, but showing the differential UVLFs. As for the SMFs, including present-day streams (Sample II, blue) and phase-mixed galaxies (Sample III, green) improves the shape and normalization of the reconstructed UVLF 
compared to intact dwarf galaxies alone (Sample I, magenta).} 
\label{fig:6}
\end{figure*}

In this section, we explore how accurately we can recover the SMFs/UVLFs of the proto-MW/LG at $z \gtrsim 6$ by adding the progenitors of present-day disrupted galaxies along with the intact galaxies in the \emph{all-progenitors} scenario. Here we divide our present-day substructures into three samples: 
\begin{description}
    \item[Sample I] Intact galaxies only (satellite and field galaxies)
    \item[Sample II] Intact galaxies and stellar streams
    \item[Sample III] Intact galaxies, stellar streams, and phase-mixed galaxies.
\end{description}
Sample II is likely more attainable than Sample III, since stellar streams usually require less observational data to identify and reconstruct than phase-mixed galaxies.

Figures \ref{fig:5} and \ref{fig:6} (top panels) compare the differential SMFs and UVLFs at $z = 7$ for Samples I (red), II (blue), and III (green) to those of the proto-MW/LG (black) and the universal SMFs/UVLFs from FIREbox$\mathrm{^{HR}}$ (yellow). As discussed in section \ref{subsec:SMFs_and_UVLFs_each_class}, we show the isolated suite (left) and the paired suite (right) separately. Solid lines show median SMFs/UVLFs from all of the simulations in the respective suite. Shaded regions show the 68\% halo-to-halo scatter except for the universal FIREbox$\mathrm{^{HR}}$ curve (yellow); here the shading denotes the 68\% confidence interval (CI) obtained via bootstrapping \citep{Feldmann_etal_2024}. We also compare the ratio of the SMF/UVLF from each sample with respect to the SMF/UVLF of the proto-MW/LG (Figures \ref{fig:5} and \ref{fig:6}, bottom panels) to help visualize their shapes as a function of stellar-mass/UV-magnitude. We compute the ratios for each simulation first and then take the median and 68\% CI. 

The proto-MW/LG SMFs/UVLFs have shapes similar to the universal SMF/UVLF from FIREbox$\mathrm{^{HR}}$, lending confidence to the assumption that the proto-MW/LG provides a representative sample of the global-average SMF/UVLF during reionization. However, the amplitude of the proto-MW/LG SMF/UVLF is lower than that of FIREbox$\mathrm{^{HR}}$, particularly in the isolated suite. This could be attributed to the difference in the calculation of the volume that goes into the denominator while computing the SMFs/UVLFs (see section \ref{sec:SMFs_and_UVLFs}). The stellar masses and UV magnitudes in FIREbox$\mathrm{^{HR}}$ are also computed differently than in this work (see \textbf{\citealt{Feldmann_etal_2024}} for details). Additionally, the FIREbox$\mathrm{^{HR}}$ box has a higher mean matter density (exactly matching the mean density of the Universe) across the simulation volume than the MW-like zoomed simulations at high redshifts, which could also lead to this difference in amplitude. This discrepancy is smaller in the case of the paired LG-like suite because these simulations have a higher mean matter density than the isolated MW-like simulations, and the galaxies form stars at earlier times in the paired suite, leading to more star-forming galaxies at high redshifts. We plan to explore this difference 
further in future work.

\begin{figure*}[ht!]
\centering
\includegraphics[width=\textwidth]{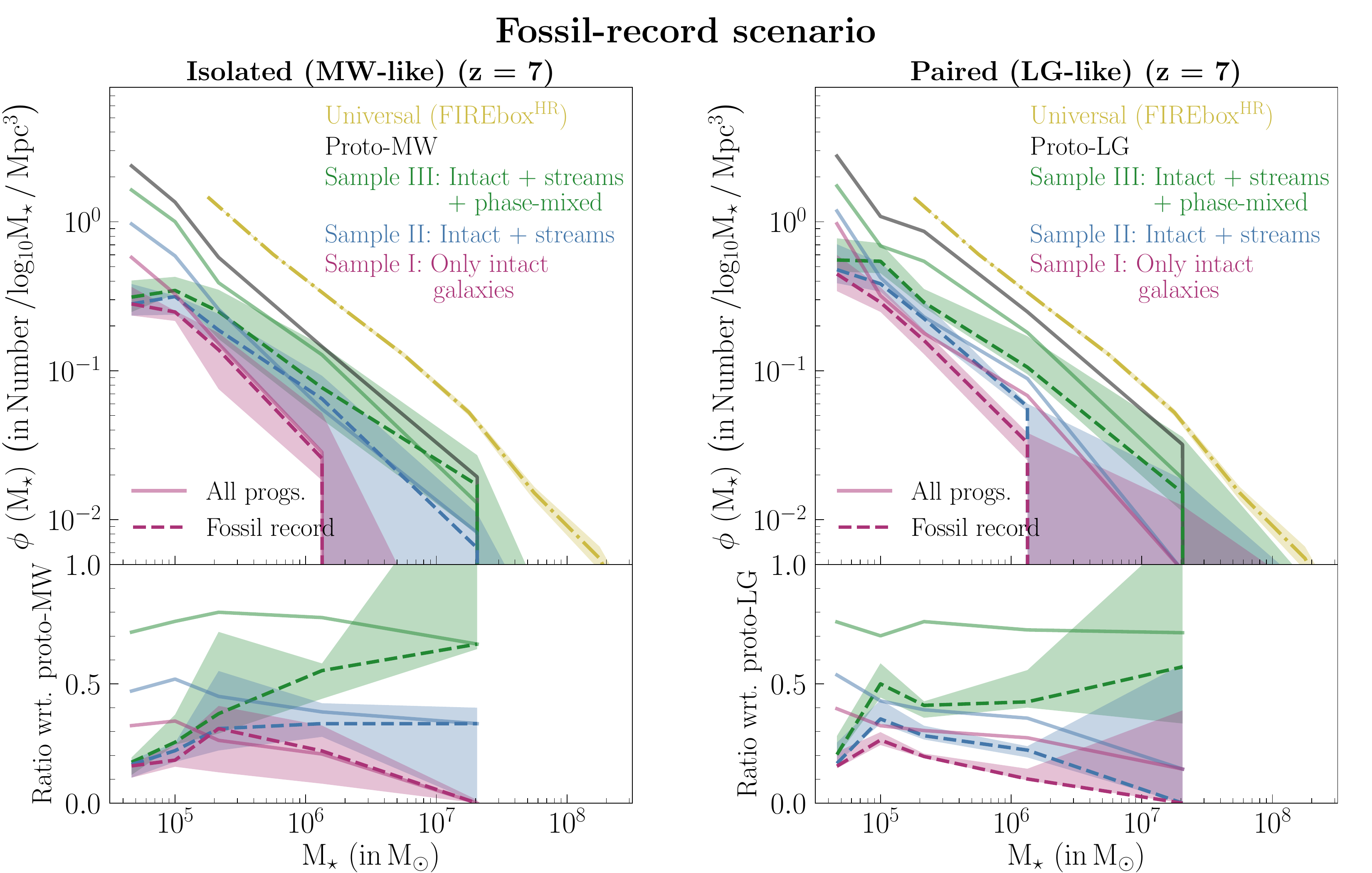}
\caption{\textbf{Differential SMFs at $z = 7$ in the \emph{fossil-record} scenario.} Similar to Figure \ref{fig:5}, but for the SMFs in the \emph{fossil-record} scenario (curves with shading). This scenario assumes all the stellar mass from multiple progenitors in enclosed in a single one, as we would infer observationally. 
The SMFs and ratios for the \textit{all-progenitors} scenario shown in Figure \ref{fig:5} are included for comparison (solid curves with no shading). 
Unlike the \emph{all-progenitors} scenario, the \emph{fossil-record} scenario has a mass-dependent bias: the reconstruction is less complete for lower-mass galaxies than for higher-mass galaxies. The effect is most prominent in Sample III (green), followed by Samples II (blue) and I (magenta) respectively.
} 
\label{fig:7}
\end{figure*}

\begin{figure*}[ht!]
\centering
\includegraphics[width=\textwidth]{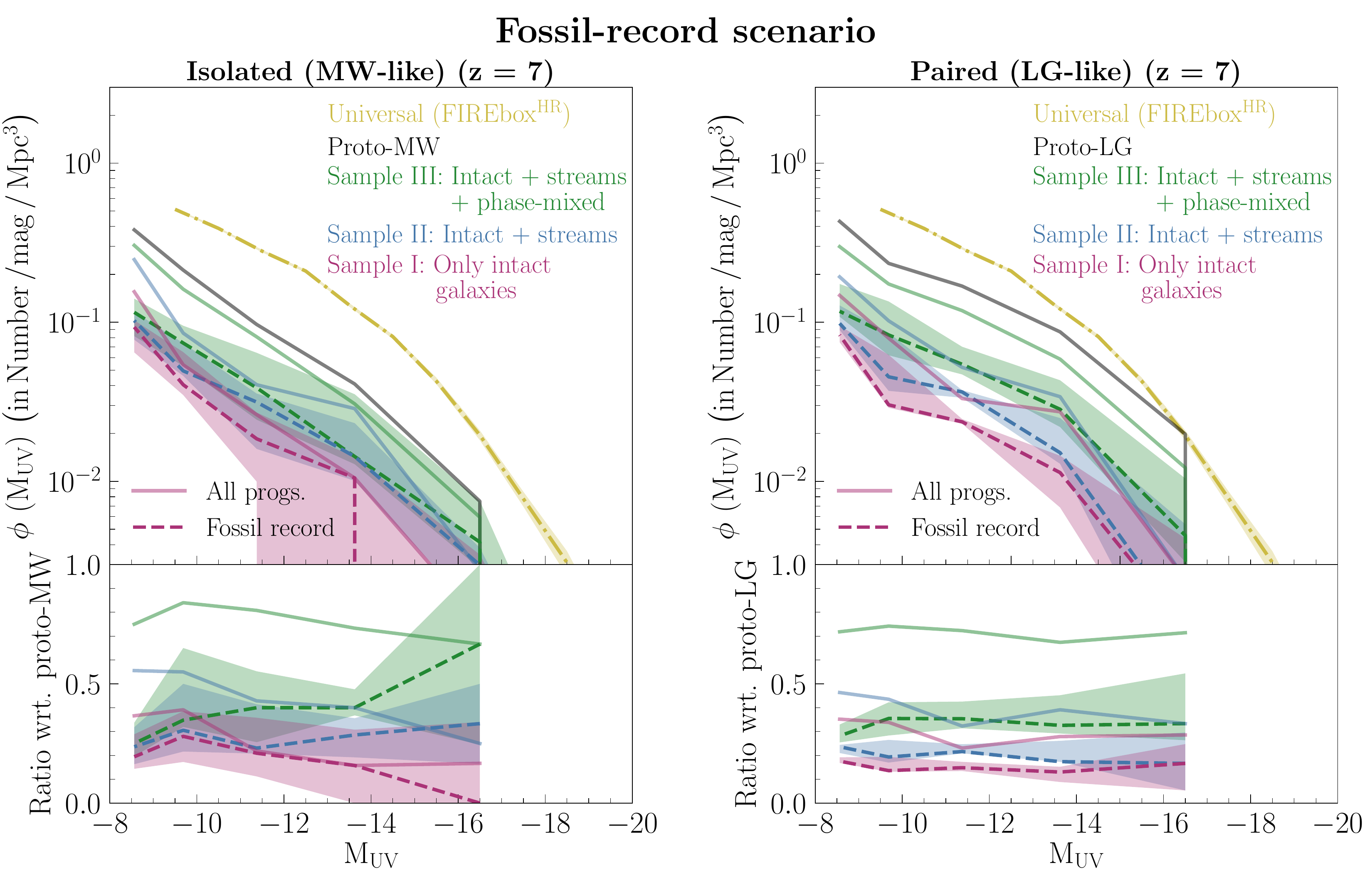}
\caption{\textbf{Differential UVLFs at $z = 7$ in the \emph{fossil-record} scenario.} Similar to Figure \ref{fig:7}, but for the differential UVLFs. The mass-dependence of the reconstruction noted in Figure \ref{fig:7} for the SMFs is less pronounced for the UVLFs, especially in the LG-like systems.
} 
\label{fig:8}
\end{figure*}

The overall shapes of the reconstructed SMFs/UVLFs from the three samples are qualitatively similar to that of the proto-MW/LG. Adding the progenitors of the stellar streams recovers the SMF/UVLF of the proto-MW/LG more accurately, particularly in terms of normalization, up to $\sim50$\%. Adding the progenitors of the phase-mixed galaxies improves the reconstruction even more. Including the disrupted systems helps most in recovering intermediate- and high-mass progenitors that are missed when using only intact low-mass galaxies. 
Furthermore, the halo-to-halo scatter in Samples II and III generally appears lower than sample I, with the exception of bottom right panels in Figures \ref{fig:5} and \ref{fig:6}. But there the scatter is from just 3 paired halos and hence may not be robust. This is even more apparent when the slopes at the low-mass/faint end is estimated (Section \ref{subsec:low-mass-/faint-end_slopes}). Adding more disrupted galaxies to our present-day samples leads to more robust recovery of our proto-MW/LG SMFs/UVLFs by potentially reducing the halo-to-halo scatter, which is sensitive to the variations in accretion histories. This is particularly prominent in the case of the isolated suite, since it probes a wider range of accretion histories than the paired suite, leading to more scatter in the case of Sample I. 

The ratios of the reconstructions from each sample to the proto MW/LG are nearly flat over a wide range of stellar masses/magnitudes. The scatter is larger in the high-mass end because of low-number statistics. Samples I and II have a slight bias at lower (fainter) masses (magnitudes), but remain almost flat at the low-mass/faint end ($M_\star \lesssim 10^{6.5} \, M_\odot$ or $M_{\mathrm{UV}} \gtrsim - 14$). Sample II recovers $\sim$50\% of the proto-MW/LG SMF/UVLF while Sample I recovers only $\sim$25\%, a factor of $\sim$2 improvement in the normalization.  Sample III does even better, recovering $\sim$75\% of the proto-MW/LG with reduced bias. These results argue that the progenitors of the stellar streams and phase-mixed galaxies are crucial for a more accurate and precise representation of the proto-MW/LG and the universal SMFs/UVLFs at high redshifts.

\subsection{Reconstructed Differential Stellar Mass Functions and Ultraviolet Luminosity Functions from the Fossil Record of Present-day Substructures} \label{subsec:SMFs_and_UVLFs_fossils}

We next reconstruct the SMFs from the fossil record, by combining the stellar masses of all the progenitors of each galaxy into a single progenitor (Figure \ref{fig:7}). We use this combined stellar mass to calculate UV magnitudes and reconstruct the UVLFs (Figure \ref{fig:8}). The color scheme in these figures remains the same as in Figure \ref{fig:5}, but the dashed lines now show the median SMFs/UVLFs from the fossil record, and the shaded regions show the 68\% halo-to-halo scatter in the fossil record. We show the SMFs/UVLFs from the \textit{all-progenitors} scenario as solid lines for reference but (for the sake of readability) do not show their halo-to-halo scatter.  

Using the fossil record instead of the complete merger history has several effects on reconstructing the $z = 7$ SMFs/UVLFs. First, the overall normalization is reduced because there are fewer progenitors in the \emph{fossil-record} scenario. Second, the SMFs/UVLFs from the \emph{fossil-record} scenario in Figures \ref{fig:7} and \ref{fig:8} (top panel) and their ratios with respect to that of the proto-MW/LG (bottom panel) decrease towards the low-mass/faint end because many of the low-mass/faint progenitors are combined into a single more massive progenitor in the fossil-record assumption. 
Since the present-day disrupted galaxies can be more massive than intact galaxies, and thus can have multiple progenitors at high redshifts, the fossil record approach leads to larger differences in shape 
towards the low-mass/faint end when they are included. Furthermore, the number of massive systems with multiple high-$z$ progenitors also increases when we add more types of present-day substructures. Hence the SMFs/UVLFs of Sample III, which contains the most substructures in general as well as the largest number of disrupted galaxies, show the largest difference, followed by Sample II and I, respectively. 

Except for the case of the Sample III SMF in the isolated suite, the ratios in the \emph{fossil-record} scenario remain nearly flat between $10^{5} \, M_\odot \lesssim M_\star \lesssim 10^{6.5} \, M_\odot$ for the other SMFs. In other words, the bias is not mass dependent over this range. The UVLFs between $-14 \lesssim M_{\mathrm{UV}} \lesssim - 10$ also remain nearly flat. However, over the entire mass/UV-magnitude range, the ratios decrease towards the low-mass/faint end. This occurs because the \emph{fossil-record} scenario combines multiple progenitors into a single effective progenitor, which reduces the number of low-mass/faint progenitors. The difference is exacerbated for Sample II and Sample III which have better completeness at higher stellar mass. We also note that the relative shape of the SMFs/UVLFs from the three samples
do not change significantly in the case of the proto-LG, particularly for the UVLFs. These results do not rule out the possibility of using disrupted galaxies along with intact galaxies in the near-far technique, but rather indicate that our inferences are sensitive to the selection criteria for \emph{both} the present-day substructures and their high-$z$ progenitors, along with the effects of the environment they reside in. They are also sensitive to whether we use the SMF or the UVLF. We explore this selection-dependence more quantitatively in the next section.

\begin{figure*}[ht!]
\centering
\includegraphics[width=\textwidth]{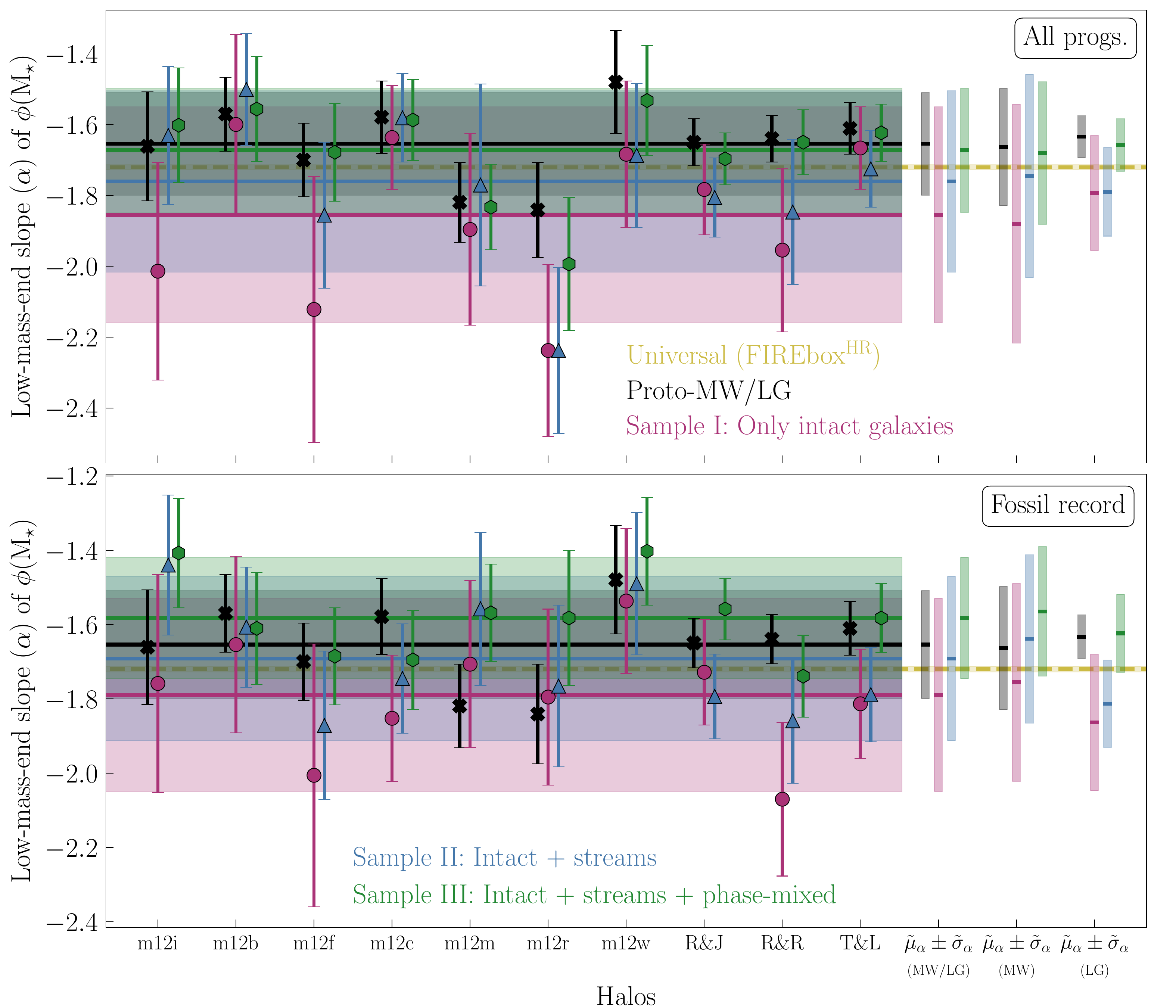}
\caption{\textbf{Adding streams and phase-mixed galaxies improves the accuracy and precision of the recovered slope of the differential SMF at low masses.}
We compare the low-mass slopes of the differential SMFs at $z = 7$ for Samples I (magenta), II (blue), and III (green) to both the full SMF of the proto-MW/LG (black) and the universal slope estimate from FIREbox$\mathrm{^{HR}}$ (yellow). Errorbars indicate the standard deviation on the recovered slope for each individual halo, while the horizontal lines show the median ($\tilde\mu_\alpha$) of the bootstrapped sample of mean $\alpha$ values across all the halos, computed as described in Section \ref{subsec:low-mass-/faint-end_slopes} and Appendix \ref{append_sec:bootstrap}.
The shaded regions show the median ($\tilde\sigma_\alpha$) of the bootstrapped sample of standard deviations in $\alpha$, representing the halo-to-halo scatter. We show ($\tilde\mu_\alpha$) and ($\tilde\sigma_\alpha$) in the rightmost points for proto-MW and proto-LG by combining them as well as separately. In the \textit{all-progenitors} scenario (top), Sample III consistently reproduces the low-mass-end slope of the proto-MW/LG SMF most accurately ($\alpha$ closest to that of the proto-MW/LG for each individual halo) and precisely (smaller errorbars for each individual halo). The median slope ($\tilde{\mu}_\alpha$) for Sample III is also closest to that of the proto-MW/LG and have the smallest halo-to-halo scatter ($\tilde{\sigma}_\alpha$), followed by Sample II and I. 
In the \emph{fossil-record} scenario (bottom), Samples II and III still have consistently smaller uncertainties for each individual halo but Sample I is more accurate than II and III at recovering the slope in some of the halos. 
The conclusions remain similar when MW and LG are separately considered.
}
\label{fig:9}
\end{figure*}

\begin{figure*}[ht!]
\centering
\includegraphics[width=\textwidth]{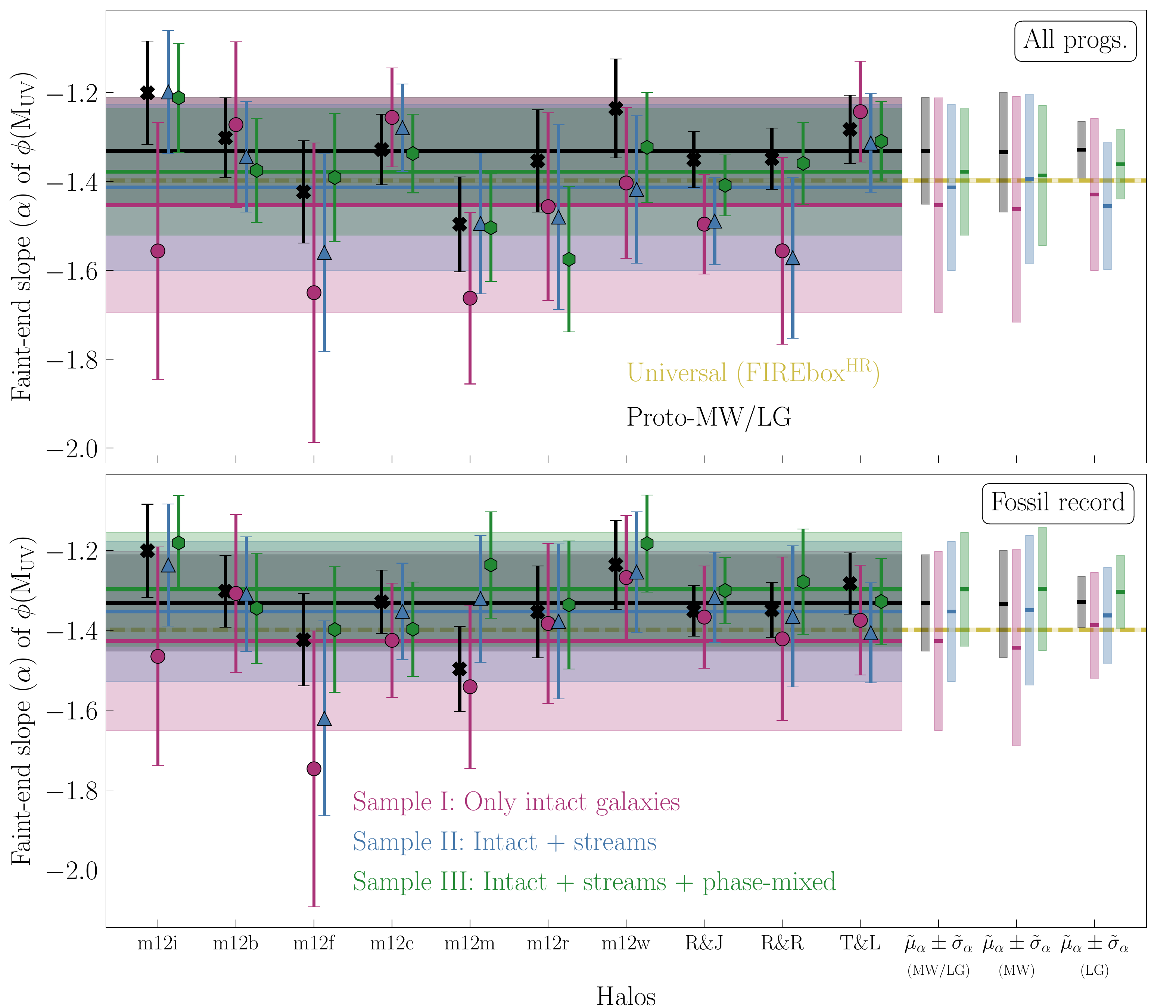}
\caption{\textbf{Adding streams and phase-mixed galaxies improves the accuracy and precision of the recovered slope of the differential UVLF at low masses} Similar to Figure \ref{fig:9}, but for the UVLFs. Overall, the UVLF slopes are shallower than the SMF slopes owing to the superlinear dependence of the UV luminosity on stellar mass (see Appendix \ref{append_sec:UV-M_star_relation}).} 
\label{fig:10}
\end{figure*}

\section{Shapes of the SMFs and UVLFs during the EoR} \label{sec:shape_SMFs_UVLFs}

\subsection{Slopes at the Low-mass/Faint end} \label{subsec:low-mass-/faint-end_slopes}

Since a primary goal of the near-far technique is to determine the low-mass-end slope of the SMF at $z \gtrsim$ 6, we next consider the effect of including the progenitors of the disrupted galaxies on the recovery of this slope. \citet{Gandhi_etal_2024} showed that the low-mass-end slope can be accurately recovered from the fossil record progenitors of just the intact low-mass galaxies, without taking disrupted galaxies into account. They compute the low-mass-end slope by fitting a single power law to the SMF across a mass range of $M_\star = 10^{4.5-6.5}$ $M_\odot$ and report the median slope across all of the simulations, concluding that the slope inferred from the fossil record of just the intact low-mass galaxies agrees with that of the proto-MW/LG. We test this result by calculating the slopes of the reconstructed high-$z$ SMFs/UVLFs for our three samples and the proto-MW/LG, but fit a Schechter function to incorporate massive galaxies as well \citep{Schechter_1976}. This is well-motivated by observations \citep{Duncan_etal_2014, Bouwens_etal_2015, Finkelstein_etal_2015, Song_etal_2016, Atek_etal_2018, Ishigaki_etal_2018, Bhatawdekar_etal_2019, Bouwens_etal_2021, Bouwens_etal_2023, Whitler_etal_2025}, but alternative models (e.g., double power law) also exist, particularly for explaining the excess number of galaxies at the bright end ($\mathrm{M_{UV} \lesssim -22})$ \citep{Bowler_etal_2015, Bowler_etal_2017, Bowler_etal_2020, Donnan_etal_2023, Harikane_etal_2023, Sun_etal_2023b, Harikane_etal_2024, Donnan_etal_2024, Chemerynska_etal_2025}, which is beyond the brightest set of galaxies in our samples. Instead of fitting the Schechter functions to binned SMFs/UVLFs, we perform Bayesian modeling using the \texttt{PyMC} package \citep{PyMC} to estimate the posterior distribution of the slope parameter $\alpha$. This approach eliminates the need for binning, which can affect the result for $\alpha$, and provides a full posterior distribution for $\alpha$. In the case of FIREbox$\mathrm{^{HR}}$ we do use the binned SMFs/UVLFs, but generate posteriors of $\alpha$ by fitting Schechter functions to these SMFs/UVLFs using the same approach. The details of our methodology, along with illustrative results, are described in Appendix \ref{append_sec:bayes_SMF_UVLF}.

We estimate the posterior distribution of the slope $\alpha$, which follows an approximately Gaussian distribution (see Appendix \ref{append_sec:bayes_SMF_UVLF}), and calculate its mean and standard deviation for the three samples in each simulation. Figures \ref{fig:9} and \ref{fig:10} show the slopes obtained by our fits for the SMFs and UVLFs respectively, compared to the ones from the proto-MW/LG and the universal SMFs/UVLFs at $z = 7$. The \textit{all-progenitors} scenario is shown in the top panel, while the \emph{fossil-record} scenario is in the bottom panel. The points and the error bars in each figure denote the mean and standard deviation of the posterior distribution for the low-mass-end/faint-end slope $\alpha$ for the proto-MW/LG (black), Sample I (red), Sample II (blue) and Sample III (green) for each simulation. The mean slope of the universal SMF/UVLF is shown by the yellow line, while the shaded region shows the standard deviation. To estimate population-level parameters, we bootstrap the 
individual $\alpha$ values, resampling from their Gaussian posteriors. This yields: (1) a central slope estimate ($\tilde{\mu}_\alpha$) - median of the bootstrapped means; (2) halo-to-halo scatter ($\tilde{\sigma}_\alpha$) - median of the bootstrapped standard deviation. We show these by the horizontal lines and the shaded regions, respectively, in Figures \ref{fig:9} and \ref{fig:10}. To compare the results between proto-MW and proto-LG, we separately do the analysis for the isolated and paired galaxy suites. For clarity, we also show these estimates in the rightmost points of both the figures. We emphasize that $\tilde{\sigma}_\alpha$ is \emph{not} the standard deviation of the bootstrapped mean $\alpha$ distribution, but rather an average range of possible $\alpha$ values across all the 
halos in the respective galaxy suite, defining the halo-to-halo scatter. We discuss the details of our bootstrapping method in Appendix \ref{append_sec:bootstrap}. Tables \ref{tab:slopes} and \ref{tab:slopes2} summarizes our slope estimates. We note that the median values for the isolated and paired galaxy suites calculated separately, in general, provide similar conclusions as the median values for the 10 halos combined. However, the results for the paired galaxy suite are derived from only three simulations and thus may not be robust. Since the trend is similar we discuss our results for $\tilde{\mu}_\alpha$ and $\tilde{\sigma}_\alpha$ when all the halos are combined, unless stated otherwise.

The figures show slope estimates using progenitor galaxies with $M_\star \geq 10^{5} \, M_\odot$ or $M_{UV} \leq -10$ (our fiducial choice). We defer a discussion of the effect of varying these cut-offs to Section \ref{subsec:select_criteria_slopes}. The proto-MW/LG slopes are representative of the universal slopes. There is an offset of $\approx0.07$ in the median slope values, but they agree within the median halo-to-halo scatter. This could be due to the difference in mean matter densities of the universe and the MW/LG region as discussed in Section \ref{subsec:SMFs_and_UVLFs_all_progs}. We will explore this aspect in future work. Since the proto-MW/LG slopes are representative of the universal values within the uncertainties, we mainly assess our results from the three samples against the proto-MW/LG for quantitative comparisons. 

The low-mass-end slope values of the proto-MW/LG and Sample I reported here are different than the ones in \citet{Gandhi_etal_2024} by $\approx$0.1, likely due to the difference in the slope estimation method. In addition, some of the present-day intact galaxies in \citet{Gandhi_etal_2024} are instead identified as part of our set of disrupted galaxies, so our Sample I is not identical to the set used in that paper.

We find several key features when we include the disrupted galaxy progenitors while computing the slopes at the low-mass/faint end. First, when we compare the individual low-mass-end/faint-end slopes from the SMFs/UVLFs of the three samples with the one from the proto-MW/LG, we find that in most halos the slope of the proto-MW/LG SMF/UVLF is consistently better reproduced by Sample III, followed by Samples II and I, respectively. This is also reflected in the median values, $\tilde{\mu}_\alpha$, from our bootstrapped samples. Second, including more disrupted galaxies reduces the standard deviation of $\alpha$ values from Sample I within each halo, by up to $\sim$30\% in Sample II and $\sim$50\% in Sample III. This is because the increased number of progenitors improves statistical robustness. Third, there is a scatter in the values of the slope across different halos, which implies that different accretion histories can lead to different values of the reconstructed low-mass-end/faint-end slope, leading to poorly constrained values across the realizations. However, we find that this halo-to-halo scatter can be reduced by adding the disrupted galaxies, as shown by the decrease in the median $\tilde{\sigma}_\alpha$ values (rightmost points). This reduction in the scatter, in comparison to Sample I, can be $\sim$20\% in case of Sample II and $\sim$40\% in case of Sample III. The effect of an unusual accretion history in producing a poorly constrained slope can thus be mitigated by adding the disrupted galaxies. 

We reach similar conclusions when we consider the effect of adding disrupted galaxies separately to a MW-like environment and a LG-like environment. The halo-to-halo scatter in case of a LG-like environment is smaller compared to that of a MW-like environment, but the results are derived from only three simulations and may not be robust. Additionally, the low-mass-end for Samples I and III is steeper than proto-LG because of the R\&R paired halo (see Figure \ref{fig:9}). Furthermore, the slopes are susceptible to how present-day and high-$z$ galaxies are selected, which we discuss in \ref{subsec:select_criteria_slopes} and hence the should be treated with caution.

Interestingly, the low-mass-end slopes of the SMFs (see Figure \ref{fig:9}) and the faint-end slopes of the UVLFs (see Figure \ref{fig:10}) differ. The UVLFs have a shallower slope at the faint-end ($\approx$ -1.4) compared to the low-mass-end of the SMFs ($\approx$ -1.7). Even though extrapolations of observed UVLFs predict a faint-end slope of $\approx$ -2, results from FIREbox$\mathrm{^{HR}}$ also seem to be pointing to a shallower slope (see Table \ref{tab:slopes}). Although this could depend on the specific distribution used to draw the progenitor masses and the slope estimation method, the difference between the SMF and UVLF slopes invites further investigation. 
Using the power-law nature of the SMF at the low-mass end and the mass-to-light ratios computed from our progenitor stellar halos, we find that the shallower slope in the UVLFs, can be partially explained due to the superlinear dependence of UV luminosity with stellar mass. Since young, massive O/B stars generate most of the UV radiation and lower-mass galaxies are less efficient at producing them, this leads to suppressed UVLFs at the faint end. We refer the reader to Appendix \ref{append_sec:UV-M_star_relation} for further details and figures demonstrating our analysis. 
In any case, we better reproduce the proto-MW/LG SMFs/UVLFs after including present-day disrupted galaxies. 

Another important difference can be discerned between the \emph{all-progenitors} scenario and the \emph{fossil-record} scenario: the latter case results in flatter slopes - by $\sim$0.07 on average across the three samples when both the isolated and paired galaxy suites are combined. 
As discussed in Section \ref{subsec:SMFs_and_UVLFs_fossils}, this effect is due to combining multiple progenitors into a single one in the \emph{fossil-record} scenario and it gets amplified for Samples II and III as they have more massive substructures with multiple progenitors.
The level of flattening varies depending on how both the present-day and high-$z$ progenitor galaxies are selected. 

\begin{deluxetable*}{lcccc}
\tablenum{2}
\tablecaption{\textbf{Low-mass-end and Faint-end Slope Estimates at $z = 7$, for Isolated and Paired Simulations Combined.}}
\label{tab:slopes}
\tablewidth{0pt}
\tablehead{
\colhead{Sample} & 
\multicolumn{2}{c}{SMFs} & 
\multicolumn{2}{c}{UVLFs} \\
\colhead{} & 
\colhead{} & 
\colhead{} & 
\colhead{} & 
\colhead{}
}
\startdata
Universal & \multicolumn{2}{c}{-1.72 $\pm$ 0.01} & \multicolumn{2}{c}{-1.40 $\pm$ 0.01} \\
Proto-MW/LG & \multicolumn{2}{c}{-1.65 $\pm$ 0.14} & \multicolumn{2}{c}{-1.33 $\pm $0.12} \\
\hline
\multicolumn{5}{c}{} \\[-0.5em]
\multicolumn{1}{l}{} & 
\multicolumn{1}{c}{All Progenitors} & 
\multicolumn{1}{c}{Fossil Record} & 
\multicolumn{1}{c}{All Progenitors} & 
\multicolumn{1}{c}{Fossil Record} \\
\hline
Sample I & -1.85 $\pm$ 0.30 & -1.79 $\pm$ 0.26 & -1.45 $\pm$ 0.24 & -1.43 $\pm$ 0.22\\
Sample II & -1.76 $\pm$ 0.26 & -1.69 $\pm$ 0.22 & -1.41 $\pm$ 0.19 & -1.35 $\pm$ 0.18\\
Sample III & -1.67 $\pm$ 0.18 & -1.58 $\pm$ 0.16 & -1.38 $\pm$ 0.14 & -1.30 $\pm$ 0.14\\
\enddata
\tablecomments{The values reported here for the three samples and the proto-MW/LG are derived from bootstrapping with replacement. The central estimates are the medians of $\mu_\alpha$ samples, while the associated uncertainties are the medians of $\sigma_\alpha$ samples, which we define as our halo-to-halo scatter. The slopes of the universal SMF/UVLF are estimated by fitting a Schechter function and the uncertainties associated with them are statistical uncertainties.}
\end{deluxetable*}

\begin{deluxetable*}{lcccccccc}
\tablenum{3}
\tablecaption{\textbf{Low-mass-end and Faint-end Slope Estimates at $z = 7$, for Isolated and Paired Simulations Separately.}}
\label{tab:slopes2}
\tablewidth{0pt}
\tablehead{
\colhead{Sample} & 
\multicolumn{4}{c}{SMFs} & 
\multicolumn{4}{c}{UVLFs} 
}
\startdata
Universal & \multicolumn{4}{c}{-1.72 $\pm$ 0.01} & \multicolumn{4}{c}{-1.40 $\pm$ 0.01} \\
Proto-MW & \multicolumn{4}{c}{-1.66 $\pm$ 0.17} & \multicolumn{4}{c}{-1.33 $\pm$ 0.13} \\
Proto-LG & \multicolumn{4}{c}{-1.63 $\pm$ 0.06} & \multicolumn{4}{c}{-1.33 $\pm$ 0.06} \\
\hline
& \multicolumn{2}{c}{All Progenitors} & \multicolumn{2}{c}{Fossil Record} &
  \multicolumn{2}{c}{All Progenitors} & \multicolumn{2}{c}{Fossil Record} \\
& \multicolumn{1}{c}{Proto-MW} & \multicolumn{1}{c}{Proto-LG} & \multicolumn{1}{c}{Proto-MW} & \multicolumn{1}{c}{Proto-LG} &
  \multicolumn{1}{c}{Proto-MW} & \multicolumn{1}{c}{Proto-LG} & \multicolumn{1}{c}{Proto-MW} & \multicolumn{1}{c}{Proto-LG} \\
\hline
Sample I & -1.88 $\pm$ 0.34 & -1.79 $\pm$ 0.16 & -1.76 $\pm$ 0.27 & -1.86 $\pm$ 0.18 &
           -1.46 $\pm$ 0.25 & -1.43 $\pm$ 0.17 & -1.44 $\pm$ 0.25 & -1.39 $\pm$ 0.13 \\
Sample II & -1.75 $\pm$ 0.29
 & -1.79 $\pm$ 0.13 & -1.64 $\pm$ 0.23 & -1.81 $\pm$ 0.12 &
            -1.39 $\pm$ 0.19 & -1.46 $\pm$ 0.14 & -1.35 $\pm$ 0.19 & -1.36 $\pm$ 0.12 \\
Sample III & -1.68 $\pm$ 0.20 & -1.66 $\pm$ 0.07 & -1.56 $\pm$ 0.17 & -1.62 $\pm$ 0.11 &
             -1.39 $\pm$ 0.16 & -1.36 $\pm$ 0.08 & -1.30 $\pm$ 0.15 & -1.30 $\pm$ 0.09 \\
\enddata
\tablecomments{See Table \ref{tab:slopes} note.}
\end{deluxetable*}

\subsection{Sensitivity to Stellar Mass Selection Criteria}
\label{subsec:select_criteria_slopes}
In Section \ref{subsec:SMFs_and_UVLFs_fossils}, we noted that the use of the fossil record seemed to increase the sensitivity of our reconstructions to the selection of galaxies used in the near-far technique. Here we explore this further, by testing how the $z=7$ SMF slope estimates change under several variations of our selection criteria.

\subsubsection{\texorpdfstring{Lowering the Low-mass Cutoff at $z = 7$}{Lowering the Low-mass Cutoff at z = 7}}

We select progenitor galaxies with $M_\star \geq 10^{4.5} \, M_\odot$ instead of $M_\star \geq 10^5 \, M_\odot$) at $z = 7$. This increases the number of progenitor galaxies, which naturally leads to a smaller standard deviation on $\alpha$ in each simulation, for each of the samples. In this case, Sample III produces the lowest standard deviation followed by Sample II and I, respectively. However, we get steeper mean $\alpha$ values in the \textit{all-progenitors} scenario and flatter values in the \emph{fossil-record} scenario. This leads to a larger difference between the slopes in these two scenarios when more disrupted galaxies are added, even though the halo-to-halo scatter decreases. Our estimates using bootstrapping, shown in Table \ref{tab:slopes}, demonstrate these effects. 
With a lower-mass cutoff, Sample III provides the most accurate and precise estimate of the proto-MW/LG slope in the \textit{all-progenitors} scenario followed by Samples II and I, respectively. However, in the \emph{fossil-record} scenario, Sample I provides a more accurate representation of the proto-MW/LG slope than Samples II and III. Thus reducing the mass cut-off incorporates more low-mass progenitors which leads to the steepening of the slope in the \textit{all-progenitors} scenario. However, when combined into a single progenitor in the \emph{fossil-record} scenario, we get flatter SMFs at the low-mass end. The effect gets more drastic when we add more massive present-day substructures since we increase the amplitude more towards the high-mass end than in the low-mass end. 

\subsubsection{Increasing the Present-day Stellar-Mass Cutoff for Intact Galaxies} 
We can reliably select present-day intact galaxies down to $M_\star \gtrsim 10^{4.5} \, M_\odot$, but disrupted galaxies only down to $M_\star \gtrsim 10^{6} \, M_\odot$. So we test our results by also selecting intact present-day galaxies with $M_\star \gtrsim 10^{6} \, M_\odot$. When we use progenitor galaxies with $M_\star \gtrsim 10^{4.5} \, M_\odot$ at high redshifts, we find similar results for the proto-MW/LG slope in the three samples, with the least halo-to-halo scatter in case of Sample III in the \textit{all-progenitors} scenario. However, in the \emph{fossil-record} scenario all the three samples give a flatter slope than that of the proto-MW/LG. Upon increasing the cut-off to $M_\star \gtrsim 10^{5} \, M_\odot$ at high redshifts, the difference between the slopes in the \textit{all-progenitors} scenario and the \emph{fossil-record} scenario reduces, as expected. In the \textit{all-progenitors} scenario the slope of Sample III is closer to the proto-MW/LG slope ($\Delta\tilde{\mu}_\alpha \approx 0.006$) than Sample I ($\Delta\tilde{\mu}_\alpha \approx 0.145$). However, in the \emph{fossil-record} scenario, Sample I is more accurate ($\Delta\tilde{\mu}_\alpha \approx 0.007$) than Sample III ($\Delta\tilde{\mu}_\alpha \approx 0.150$).

\subsubsection{Removing Morphologically Complex and Extremely Massive Present-day Substructures}

Our present-day sample includes a few morphologically-complex, massive galaxies with $M_\star \gtrsim 10^{9} \, M_\odot$. These typically have $\sim 30-50$ progenitors and lead to more bias in the fossil-record scenario. Hence we test our results by restricting our present-day galaxy sample to those with stellar masses between $10^{6} \, M_\odot \lesssim M_\star \lesssim 10^{9} \, M_\odot$. The trends in this case resemble our previous results without a high-mass cut-off. 

\subsubsection{Lowering the High-mass Cutoff at the Present Day}
We tried lowering the high-mass cutoff for present-day substructures to $M_\star \lesssim 10^{8} \, M_\odot$ while keeping the low-mass cutoff at $M_\star \gtrsim 10^{4.5} \, M_\odot$. In this case, we find that the slopes across the three samples in the all progenitors scenario, although similar, are steeper than that of the proto-MW/LG by $\approx 0.23$.
In the \emph{fossil-record} scenario, on the other hand, the slopes are flatter than the proto-MW/LG; with Sample III being flatter (by $\approx$0.4) than Sample II (by $\approx$0.3) or Sample I (by $\approx$0.2). Interestingly, the halo-to-halo scatter reduces only by $\sim$3\% in Sample II and by $\sim$15\% in Sample III. This remains similar when selecting progenitor galaxies with $M_\star \gtrsim 10^{5} \, M_\odot$ at high redshifts, although in this case all three samples produce a steeper slope than the proto-MW/LG (by $\approx \, 0.1-0.3$; Sample III is more accurate than Sample I).

\subsubsection{Discussion of Sensitivity Analysis}

Except when decreasing the upper mass limit, these results show that adding disrupted galaxies robustly reduces the halo-to-halo scatter, with the quantitative benefit depending on the sample selections employed. However, we can get a biased estimate of the slope in the \emph{fossil-record} approach, while still reducing the halo-to-halo scatter by adding more disrupted galaxies. The bias exists even while using just the intact galaxies, with the precise bias depending, again, on sample selection. The \emph{all-progenitors} scenario provides more accurate and precise recoveries of the underlying proto-MW/LG slope when disrupted galaxies are added. This motivates future efforts to combine the traditional fossil record analysis with simulated merger trees, which may allow one to properly account for the multiple high redshift progenitors of individual present-day substructures, both intact and disrupted. If, instead, only the fossil record is used, it is important to test sensitivity to the selection criteria used for both present-day and high-$z$ galaxies.  


\subsection{Redshift Evolution} 
\label{subsec:slope_evolve}

\begin{figure*}[ht!]
\centering
\includegraphics[width=\textwidth]{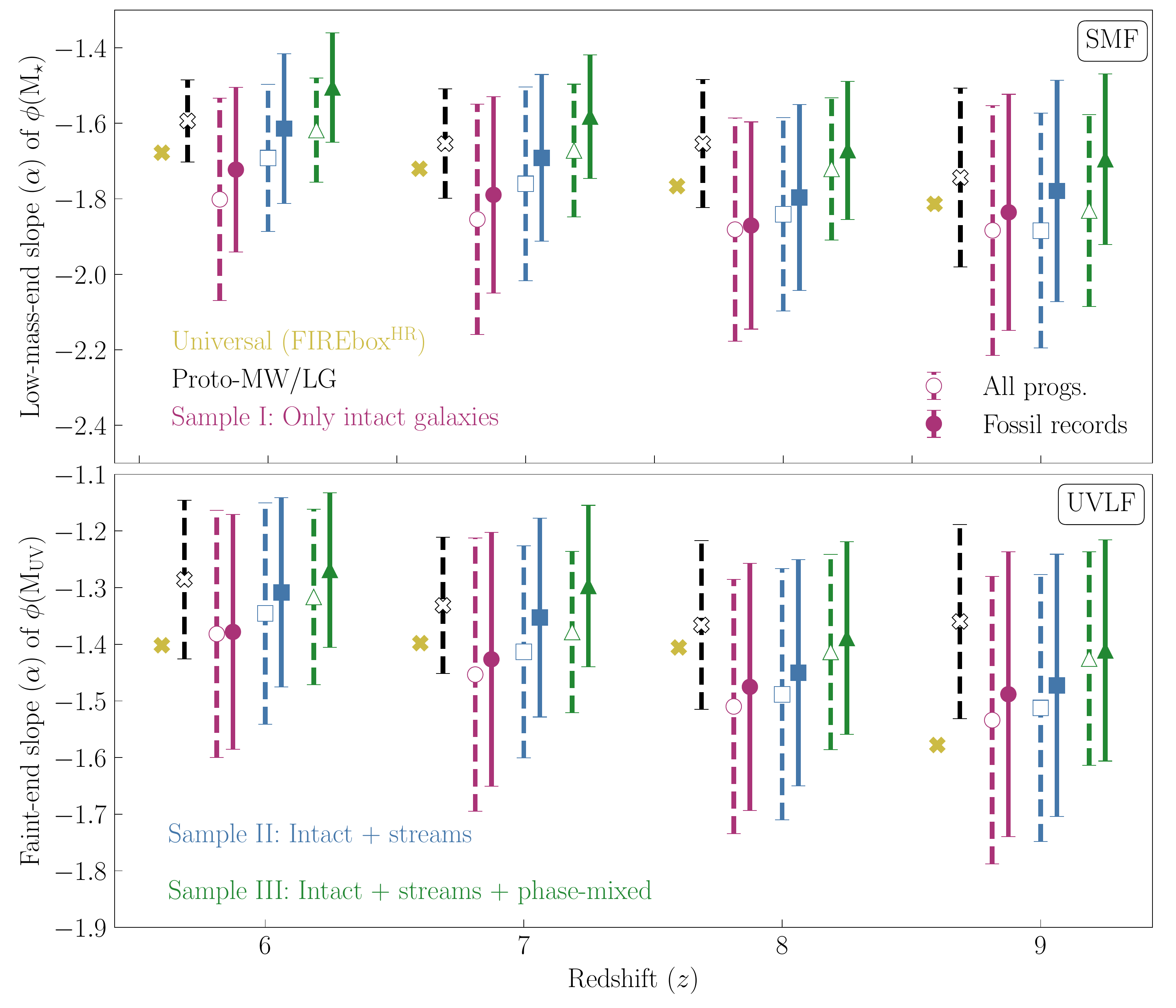}
\caption{\textbf{Redshift evolution of the low-mass-end/faint-end slope of the SMF/UVLF.} Hierarchical assembly is reflected in the increasingly steep low-mass-end/faint-end slopes with redshift for all three samples, as well as the proto-MW/LG (black) and FIREbox$\mathrm{^{HR}}$ (yellow), for the differential SMF (top) and UVLF (bottom). In both panels, black, red, blue, and green points show $\tilde{\mu}_\alpha$ values for the proto-MW/LG, Samples I, II, and III respectively;  errorbars show the $\tilde{\sigma}_\alpha$ values as a measure of halo-to-halo scatter.
Empty markers indicate slopes estimated in the \textit{all-progenitors} scenario, while filled markers denote the \emph{fossil-record} case.  
}
\label{fig:11}
\end{figure*}

We also estimate $\alpha$ for the SMFs/UVLFs at $z =$ 6, 8, and 9, using our fiducial choice of $M_\star \geq 10^{5} \, M_\odot$ or $M_{UV} \leq -10$ at high redshifts. We show the $\tilde{\mu}_\alpha$ and $\tilde{\sigma}_\alpha$ statistics as our central estimates for the low-mass-end/faint-end slope and the halo-to-halo scatter, respectively. The top panel of Figure \ref{fig:11} shows results for the low-mass-end slopes of the SMFs and the bottom panel shows the faint-end slopes for the UVLFs. The slope estimates from the \textit{all-progenitors} scenario is shown by the empty markers and dashed errorbars, while the filled markers and solid errorbars denote the slope estimates from the \emph{fossil-record} scenario. As a consequence of hierarchical structure formation in the cold dark matter picture with standard radiative and supernova feedback, the SMFs/UVLFs should generally get steeper towards higher redshifts. 
This is well-reflected by the SMFs/UVLFs from FIREbox$\mathrm{^{HR}}$ (yellow), the proto-MW/LG (black), and all the three samples (I - red; II - blue; III - green). 

In general, Sample III consistently reproduces the most accurate and precise proto-MW/LG slope across the redshifts followed by Samples II and I, in the \textit{all-progenitors} scenario. The trend is similar in the \emph{fossil-record} scenario, except in a few cases. The reduction in the halo-to-halo scatter with Samples II and III is evident across the range of redshifts shown. In particular, at $z =$ 8 and 9, the large scatter in the slope from the progenitors of the intact galaxies alone is fairly well mitigated upon adding the disrupted galaxies, as shown by Sample III. 
We caution, however, that the $z=8$ and $z=9$ results may be subject to the limited statistical samples of progenitor galaxies captured in our simulations at these redshifts.
At $z =$ 9, the faint-end slopes from our reconstructed UVLFs are shallower than the universal one. These estimates are probably poor because of the low halo abundance at these redshifts. Overall, for both the SMFs/UVLFs and for most redshifts, Sample III reproduces the proto-MW/LG slope more accurately and also reduces the halo-to-halo scatter, followed by Samples II and I. This is true for both the \textit{all-progenitors} and \emph{fossil-record} scenarios using our fiducial selection criteria. However, as discussed in section \ref{subsec:low-mass-/faint-end_slopes}, the results can vary depending on our sample selection criteria, both at the present-day and at high redshifts.

\section{Reconstructed Cosmic Ultraviolet Density} \label{sec:cosmic uv density}

\begin{figure*}[ht!]
\centering
\includegraphics[width=\textwidth]{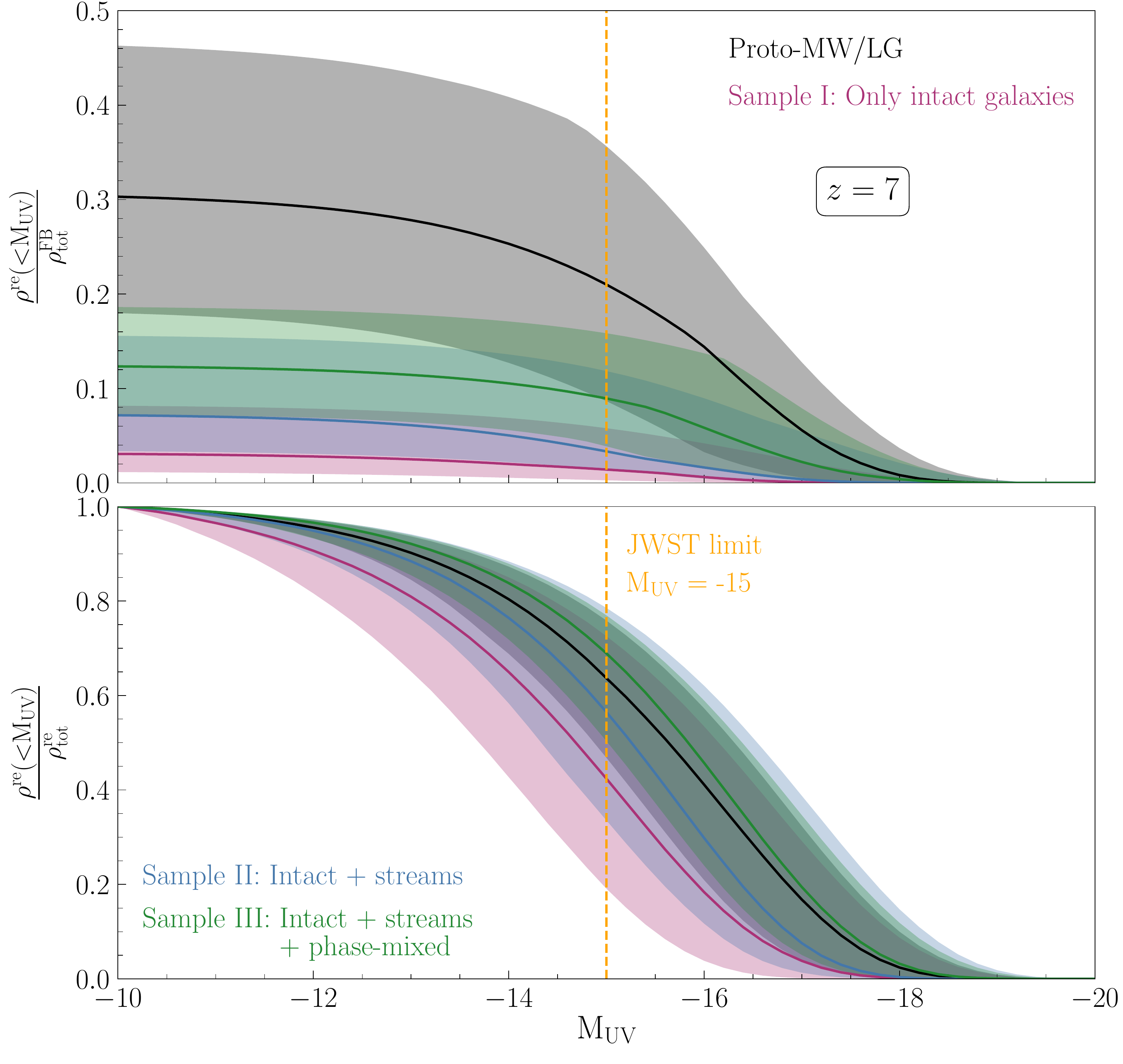}
\caption{\textbf{Reconstructed cosmic UV density at $z = 7$ using the fossil record}. \textbf{\textit{Top}}: contribution of different progenitor galaxies with UV magnitude brighter than $M_{\mathrm{UV}}$ from the fossil records of the different samples to the \emph{universal} cosmic UV density calculated from FIREbox$\mathrm{^{HR}}$. Solid lines shows the median across all of the halos; shaded regions enclose the 16th to 84th percentile range for the proto-MW/LG (black) and Samples I (red), II (blue), and III (green). The orange dotted line shows JWST's current observational threshold for unlensed galaxies at $z \sim 7$ \citep{Adams_etal_2024}. 
\textbf{\textit{Bottom}}: relative contribution of different progenitor galaxies to the total \textit{reconstructed} cosmic UV density within each sample. Including progenitors of disrupted galaxies reduces the uncertainty and predicts greater contributions from brighter galaxies, more representative of the proto-MW/LG. See Section \ref{sec:cosmic uv density} for further discussion.} 
\label{fig:12}
\end{figure*}

In this section, we reconstruct the cosmic UV density $\rho_{\rm UV}$ using the near-far technique, quantifying the contributions of different progenitor populations to the total cosmic UV density and hence their roles in reionization. 
We use the posterior distributions of the Schechter function parameters, $\alpha$, $M_{\mathrm{UV}}^{\star}$, and $\Phi^{\star}$,
from our Bayesian calculations (see Appendix \ref{append_sec:bayes_SMF_UVLF}) to construct UVLF samples. Then we compute UV densities from the UVLFs using the following equations:

\begin{equation}
    \rho_{\mathrm{UV}}^{\mathrm{re}}(>L_{\mathrm{UV}}) = \int_{L_{\mathrm{UV}}}^{\infty} L_{\mathrm{UV}} \Phi^{\mathrm{re}}(L_{\mathrm{UV}}) dL_{\mathrm{UV}}
\end{equation}

\begin{equation}
    \rho_{\mathrm{UV; tot}}^{\mathrm{re}} = \int_0^{\infty} L_{\mathrm{UV}} \Phi^{\mathrm{re}}(L_{\mathrm{UV}}) dL_{\mathrm{UV}}
\end{equation}

\begin{equation}
    \rho_{\mathrm{UV; tot}}^{\mathrm{FB}} = \int_0^{\infty} L_{\mathrm{UV}} \Phi^{\mathrm{FB}}(L_{\mathrm{UV}}) dL_{\mathrm{UV}}
\end{equation}

where $\rho_{\mathrm{UV}}^{\mathrm{re}}(>L_{\mathrm{UV}})$ (or equivalently $\rho_{\mathrm{UV}}^{\mathrm{re}}(<M_{\mathrm{UV}})$) is the reconstructed cosmic UV density obtained from our reconstructed \emph{fossil-record} UVLFs  $\Phi^{\mathrm{re}}(L_{\mathrm{UV}})$ due to the sources with UV luminosities $L_{\mathrm{UV}}$ \footnote{$M_{\mathrm{UV}}$ = 51.6 - 2.5 $\log_{10}$ ($L_{\mathrm{UV}}$/[$\mathrm{erg} \, \mathrm{s^{-1}} \, \mathrm{Hz^{-1}}$]) \citep{Oke_1974}} and brighter, $\rho_{\mathrm{UV; tot}}^{\mathrm{re}}$ is the total cosmic UV density from all the sources, and $\rho_{\mathrm{UV; tot}}^{\mathrm{FB}}$ is the total universal cosmic UV density computed by using the universal UVLF $\Phi^{\mathrm{FB}}(L_{\mathrm{UV}})$ from FIREbox$\mathrm{^{HR}}$. Our analysis is done in the \emph{fossil-record} scenario in order to mimic real-world application of the near-far technique.

For each of the three samples in each halo, we compute the reconstructed cosmic UV density and explore their contribution to the total cosmic UV density. 
As shown in Figure \ref{fig:12} (top panel), the proto-MW/LG (which also includes the progenitors of the present-day main host galaxy), can recover between $\sim$18\% and $\sim$45\% of the total universal cosmic UV density. We cannot fully recover the total universal cosmic UV density because of the difference in normalization between the UVLFs here and that of FIREbox$\mathrm{^{HR}}$, as previously discussed in section \ref{sec:SMFs_and_UVLFs}. Hence, we mainly compare the contribution from the three samples and the proto-MW/LG. Sample III ($\sim$8--18\%) is the most complete, followed by Samples II ($\sim$4--16\%), and sample I ($\sim$1--8\%).  

We also consider the relative contribution of different progenitor galaxies to the total \emph{reconstructed} cosmic UV density within each sample at $z = 7$ (Figure \ref{fig:12}, bottom panel).
The median UV luminosity density reconstructed from Sample I (red) under-predicts the contributions from galaxies brighter than $M_{\mathrm{UV}} \sim -15$, which is the current JWST detection limit for unlensed galaxies \footnote{Measurements behind foreground lensing clusters can probe fainter galaxies down to $M_{\mathrm{UV}} \gtrsim -13$ \citep{Chemerynska_etal_2025}, but they come with systematic uncertainties.}.
Furthermore, the large scatter in the faint-end slope of the reconstructed UVLF from Sample I also leads to sizable uncertainties in the inferred UV luminosity density and the relative importance of different galaxies in reionization. 
Quantitatively, as inferred from Sample I, the luminosity above which galaxies produce $\sim 50\%$ of the total UV density is somewhere in the range $-16 \lesssim M_{\mathrm{UV}} \lesssim -13.5$.  
Adding the progenitors of the disrupted galaxies in Samples II and III (blue and green shaded regions, respectively) reduces the scatter. Sample II predicts minimum UV magnitudes in the range $-16.5 \lesssim M_{\mathrm{UV}} \lesssim -14.5$, while Sample III predicts a range of $-16.5 \lesssim M_{\mathrm{UV}} \lesssim -15$. Furthermore, the median curves (solid curves) are more representative of the proto-MW/LG curve (black) when the disrupted galaxies are included. The median curve for Sample III (green) shows slightly larger contributions from brighter galaxies than in the proto-MW/LG, 
because it predicts a shallower faint-end slope than that of the proto-MW/LG in the fossil-record scenario (see Figure \ref{fig:10}).

Interestingly, the results from Samples II, III, and the proto-MW/LG in the bottom panel of Figure \ref{fig:12} indicate that contrary to the conclusion drawn from the intact galaxies alone, UV-bright progenitor galaxies ($M_{\mathrm{UV}} \lesssim -15$) contribute significantly to the total cosmic UV density. 
For example, the medians for the proto-MW/LG and Sample III predict that galaxies with $M_{\mathrm{UV}} \lesssim -15$ contribute up to $\sim$65\% and $\sim$70\% of the total cosmic UV density at $z = 7$, respectively. On the other hand, in Sample II this number is $\sim55\%$, while it is only $\sim40\%$ in Sample I. Thus, the descendants of UV-bright reionization-era galaxies may end up largely in streams and phase-mixed galaxies. An accurate census of the reionization-era UV luminosity density from the near-far technique thus requires accounting for these disrupted galaxies. 

\section{Detectability of the Present-day Substructures} \label{sec:detectability}

Our results indicate that the present-day streams and phase-mixed galaxies are important for accurately and precisely inferring the high-$z$ UVLF/SMF, so we next consider their detectability.
It is generally more challenging to detect these substructures than intact galaxies of the same total stellar mass/luminosity. 
Stellar streams from low-mass galaxies are spread across their host galaxy, resulting in low surface brightnesses ($\mu \gtrsim 27$ mag arcsec$^{-2}$) and making resolved photometric measurements, needed to reconstruct their SFHs, more challenging and time-consuming than for intact low-mass galaxies. 
Nevertheless, numerous stellar streams have been detected in the MW \citep{Helmi_etal_1999, Ibata_etal_2001, Majewski_etal_2003, Grillmair_etal_2006, Newberg_etal_2009, Shipp_etal_2018, Malhan_Ibata_2018, Li_etal_2019, Malhan_etal_2021}. Likewise, some phase-mixed galaxies have been detected in the MW, including the Gaia-Enceladus Sausage \citep{Helmi_etal_2018, Belokurov_etal_2018, Haywood_etal_2018, Myeong_etal_2018a, Myeong_etal_2018b, Johnson_etal_2023}, Sequoia \citep{Barba_etal_2019, Myeong_etal_2019}, and Thamnos \citep{Koppelman_etal_2019}, although the techniques employed are different and more complex than those used for streams. 

We expect important advances in identifying streams and phase-mixed galaxies with upcoming surveys. With LSST and Roman, we will be able to probe fainter surface brightness levels and detect more stellar streams, while also improving the quality of the existing sample. Roman will also resolve streams and satellites around nearby galaxies \citep{Lancaster_etal_2022, Pearson_etal_2022}, potentially increasing our near-far sample beyond the LG. 
Here, we estimate surface brightness sensitivities to explore the detectability of stellar streams and intact galaxies with future LSST and Roman observations around the MW and external galaxies, respectively. 
We defer a study of the detectability of phase-mixed galaxies to future work, since in general they cannot be detected with photometry alone.
We emphasize that here we strive only for preliminary surface brightness estimates, rather than a complete and accurate detectability test such like the one presented in \citet{Shipp_etal_2023} with DES. These approximate results may provide a useful precursor to more detailed mock catalogs for LSST and Roman, which we defer to future work. 

\begin{figure*}[ht!]
\centering
\includegraphics[width=\textwidth]{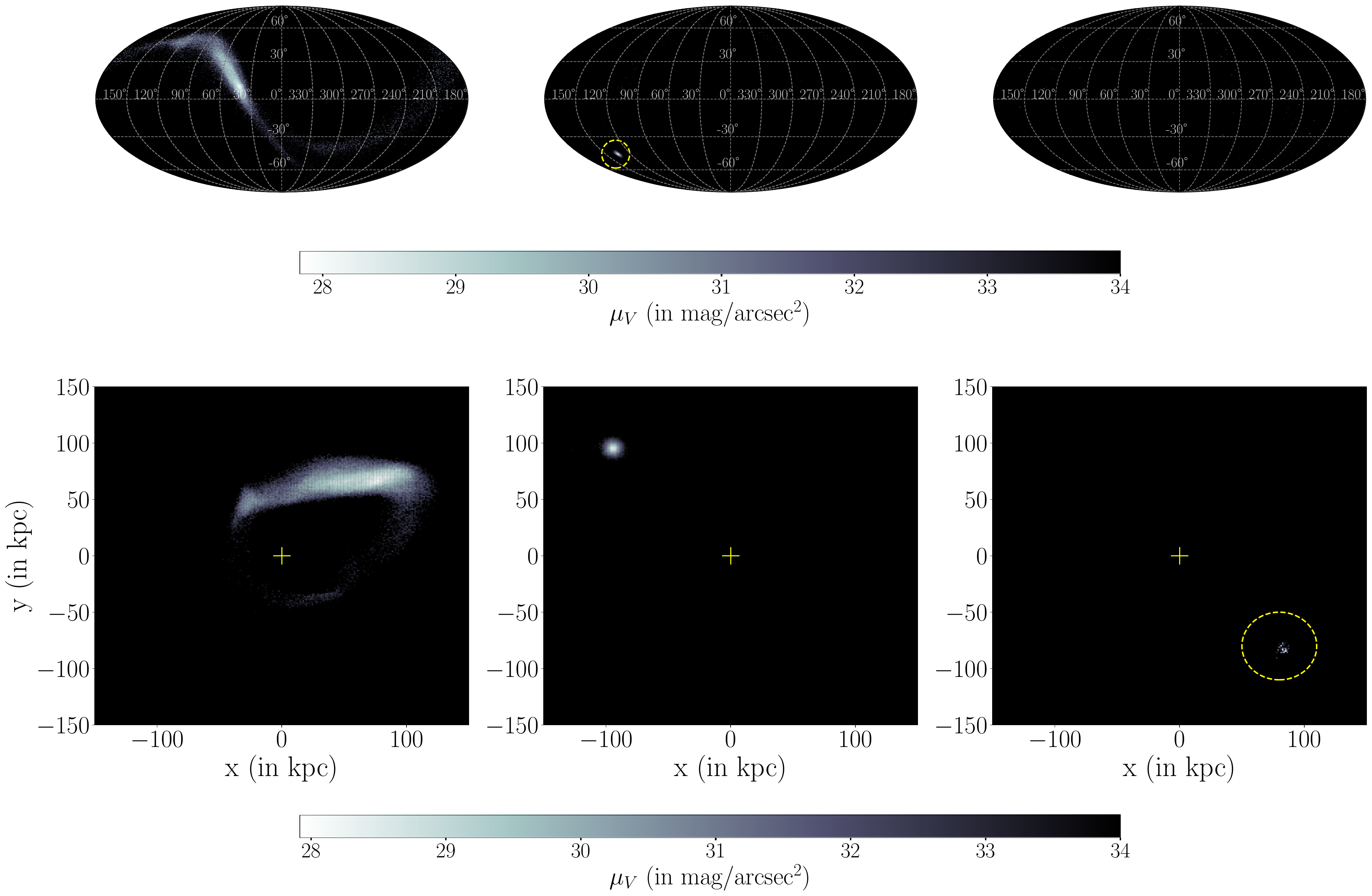}
\caption{\textbf{Detectability of streams in the MW and external galaxies}. \textbf{\textit{Top row:}} Surface brightness maps of three different streams as viewed from the center of the MW (``internal view'') using LSST, with a faint-end limit of $\mu_V$ = 34 $\mathrm{mag}$ $\mathrm{arcsec^{-2}}$. \textbf{\textit{Bottom row:}} Surface brightness maps of the same three streams, but now viewed from a distance of 5 Mpc (``external view'') to the same surface brightness limit, as expected for Roman. 
The yellow markers correspond to the center of the host galaxy. \textbf{\textit{Left:}} An easily detectable stream with long tidal tails in both internal and external views. \textbf{\textit{Center:}} a stream that would be classified as an intact low-mass galaxy in both views because of its faint tidal tails. \textbf{\textit{Right:}} a stream that is unlikely to be detectable in the internal view, but would be detected as a faint low-mass galaxy in the external view (inside yellow circle). This example was chosen to show a case which is almost visible from the internal view. 
} 
\label{fig:13}
\end{figure*}

We calculate the luminosity of each star particle in the V-band, assuming a mass-to-light ratio of $M_\star/L_V = 2$ \citep{Bullock_Johnston_2005}, and use this to calculate the surface brightness of each stream and intact galaxy in both internal and external views. For the ``internal view'' given by the LSST of MW streams (Figure \ref{fig:13}, top row), we put the observer at the center of the host galaxy and use the Galactocentric distance of each star particle to calculate their respective fluxes. We use \texttt{HEALPY} to divide the sky into pixels of equal solid angle and calculate the surface brightness by dividing the total flux in each pixel by its solid angle. We divide the simulated sky into 256 equal areas and smooth the resulting surface brightness maps with a Gaussian kernel of $\sigma=0.3 \mathrm{deg}$, which sets the angular resolution of our final ``image''. 
These values lead to a sufficient flux in each illuminated pixel to discern surface brightness variations without making the image too coarse. We choose a surface brightness limit of 34 $\mathrm{mag} \,\mathrm{arcsec}^{-2}$ as our detectability threshold. 

Among 13 simulated systems, an average of $\sim$29\% of streams should be detectable with visible, extended tidal tails by LSST,  while an additional $\sim$29\% of streams should be detectable without the tidal tails as intact low-mass galaxies. \citet{Shipp_etal_2023} performed a similar analysis using DES-detectable streams in FIRE, but excluded the massive streams ($M_\star \sim 5 \times 10^9,M_\odot$) included in our work. The detectability fraction for streams with extended tidal tails is comparable between the two studies, while our fraction for intact low-mass galaxies is slightly lower, likely due to differences in the stream samples considered.
Although we do not show examples of intact galaxies here, we estimate detectability rates of $\sim$35\% for satellites and $\sim$18\% for field galaxies. Adding the streams detectable as intact low-mass galaxies increases the percentage to $\sim$43\% for satellites.
These results illustrate the general promise of LSST for detecting intact and disrupted galaxies, but are only a first step towards more detailed investigations.  For example, the angular resolution of LSST is actually significantly higher than in these preliminary calculations, while our V-band magnitude estimates lie in between the actual LSST $u$ and $g$ bands.

For analyzing the detectability of streams and satellites of external galaxies using Roman, we employ the same method to calculate the luminosities of the simulated star particles. We then place the observer at a distance of 5 Mpc from the center of the galaxy, and calculate the flux and surface brightness in 50 arcsecond angular pixels. This angular scale spans a physical size that is larger than the mean inter-particle separation in the simulations. The bottom panel of Figure \ref{fig:13} shows these results, for the same set of three streams and a surface brightness limit
of 34 $\mathrm{mag} \, \mathrm{arcsec}^{-2}$ (based on the depth reached by the High-Latitude Wide Area Survey for resolved stars). 
As for the internal view, the stream in the left panel should be easily detectable and the stream in the middle panel may be detected as an intact low-mass galaxy. However, although the stream in the right panel is undetectable in the internal view, it passes our surface brightness threshold in the external view and is detectable as a very faint intact galaxy. We find similar results for a distance of 20 Mpc, although the resulting images are less sharp. Overall, we find that $\sim$12 \% of the streams have tidal tails detectable by Roman at this surface brightness, while $\sim$39 \% would be classified as intact low-mass galaxies. The detectability percentages for simulated satellites and field galaxies are $\sim$29 \% and $\sim$17 \%, respectively.

These results support the notion that significant populations of stellar streams and intact galaxies may be detectable in the MW and external galaxies with the LSST and Roman, and could potentially be used to probe the low-mass/faint-end of the SMF/UVLF during the EoR. In future work, however, it will be important to refine our detectability estimates with more realistic LSST and Roman synthetic observations. 

Our analysis reveals that some of the streams could be misidentified as intact low-mass galaxies, which could (incorrectly) imply that high-$z$ progenitor galaxies are less disrupted than in our simulations. This, in turn, might be used to mistakenly infer smaller contributions of disrupted progenitors to the reionization-era stellar mass budget. 
Furthermore, the SFHs derived from these present-day substructures will be incomplete until deeper observations are carried out.
Therefore, caution must be taken when reconstructing SFHs in the near-far technique, as these effects can influence the inferred shape of the SMFs/UVLFs. In our analysis here, note that we assign all galaxies with detectable streams in the streams category, rather than including sub-classifications related to the presence of the tidal tails. 

To complete our analysis, we construct UVLFs by imposing the detectability constraints obtained from the internal view scenario using LSST. Figure \ref{fig:4}---first shown in Section \ref{subsec:SMFs_and_UVLFs_each_class}---gives the differential UVLFs at $z=7$ reconstructed from different classification groups (solid curves) in comparison to reconstructions from their detectable counterparts alone (dashed curves). The reconstructed UVLFs appear promising, even after imposing the detectability constraints. Although the normalization of the satellite and field galaxy progenitor UVLFs is reduced, their shapes are relatively well-preserved. The progenitor UVLFs reconstructed from the streams remain almost unchanged even with just $\sim$50 \% detectability, since the 
detectable streams are mostly of higher mass and each massive stream typically has numerous reionization-era progenitors (see Section \ref{subsec:galaxy_assembly}), spanning much of the high-$z$ distribution in progenitor luminosity. 
We show the results in the \emph{all-progenitors} scenario by utilizing the full merger histories of the present-day substructures, and not in the \emph{fossil-record} scenario. However, the difference between the shape of the UVLFs with and without detectability constraints does not change significantly in the \emph{fossil-record} scenario.

\section{Summary} \label{sec:summary}

In this work, we test whether including present-day disrupted galaxies like stellar streams and phase-mixed galaxies in the near-far technique improves the inference of the low-mass-end/faint-end slope of the SMF/UVLF during the EoR. Using the FIRE-2 cosmological zoom-in baryonic simulations of MW-like (isolated) and LG-like (paired) halos (Section \ref{sec:simulations}), we classify present-day substructures into 4 categories: intact satellite galaxies, field galaxies, stellar streams, and phase-mixed galaxies (Section \ref{sec:substuctures_and_progs}). The first two categories constitute our intact galaxies sample (Sample I); the intact galaxies and streams comprise our Sample II; adding the phase-mixed galaxies makes our Sample III. Using particle tracking, we assign each star particle to its progenitor galaxy at high redshifts. We then assess the contribution of different progenitor populations to the total UV luminosity density (Sections \ref{sec:mass_budget_and_struct_form}--\ref{sec:shape_SMFs_UVLFs}). Finally, we estimate the proportion of substructures that will be detected by LSST (for the Milky Way) or Roman (for external galaxies) and the effect of these detection limits on the reconstructions (Section \ref{sec:detectability}). Our results are summarized below.

\textit{\textbf{Our results are consistent in general with \citet{Gandhi_etal_2024}, with a few differences.}}
\citet{Gandhi_etal_2024} estimated the low-mass-end slopes of the reconstructed fossil record SMFs using just intact low-mass galaxies and found close agreement with the proto-MW/LG slopes. We reach similar conclusions but also show that the halo-to-halo scatter associated with the slope is large when using just intact galaxies. Their slope estimates are not exactly the same as our estimates for Sample I, since some of their present-day low-mass galaxies are part of our streams and hence do not show up in our Sample I. This also leads to differences in the shape and normalization of our respective intact low-mass galaxy fossil-record SMFs, with a stronger mass dependent bias in our case. Additionally, our methods for computing the slopes are different. Furthermore, Gandhi et al. (in prep.), use an alternative set of zoom-in high-redshift FIRE-2 simulations as a benchmark for the universal SMFs/UVLFs, instead of FIREbox$\mathrm{^{HR}}$. Both sets of simulations show good agreement with each other, as well as with observed high-$z$ SMFs/UVLFs and cosmic UV luminosity density, but they may not match exactly.

\textit{\textbf{The progenitors of streams and phase-mixed galaxies represent a larger fraction of the stellar mass in the proto-MW/LG at $z = 6-9$ than the intact galaxies commonly used for the near-far technique.}}
Phase-mixed galaxy progenitors make the dominant contribution to the reionization-era population of low-mass galaxies in both the isolated and paired galaxy simulation suites. Thus, reionization in the proto-MW/LG may have been primarily driven by galaxies which subsequently merged/disrupted into their present-day host galaxies. In the isolated simulations, the second largest contribution comes from the stream progenitors, followed by the field galaxies. In the paired simulations, the field galaxies play a larger role than the streams, due the larger abundance of field galaxies in the paired simulation suite (likely an effect of the larger mean density and earlier formation time). In both simulation suites, the surviving satellite galaxy progenitors produce the \emph{smallest} fraction of the $z \sim 6-9$ stellar mass. Ignoring the progenitors of disrupted galaxies while using the near-far technique can thus lead to less accurate 
conclusions regarding the low-mass-/faint-end part of the high-$z$ SMFs/UVLFs, due to an incomplete progenitor population.

\textit{\textbf{Inclusion of tidally disrupted galaxies leads to more accurate high-$z$ reconstructed SMFs/UVLFs.}} We analyze the SMFs/UVLFs under two scenarios: the \textit{all-progenitors} scenario, in which we account for the stellar mass contribution from each progenitor of a present-day substructure separately; and the \emph{fossil-record} scenario, where we assume that the stellar mass from all progenitors of each substructure is contained within a single galaxy, as would be inferred observationally in the absence of merger history information. In both of these scenarios, our Sample III recovers most of the amplitude ($\sim$75\%) of the proto-MW/LG SMF/UVLF, followed by Samples II ($\sim$50\%) and I ($\sim$25\%), respectively. This represents a significant improvement, reducing the normalization rescaling from a factor of $\sim$4 in Sample I to just $\sim$1.5 in Sample III, after including the disrupted galaxies. However, in the \emph{fossil-record} scenario, the SMFs/UVLFs become shallower in the low-mass-end/faint-end region, with the effect being large when more disrupted galaxies are added. Nevertheless, in most instances, the SMFs and UVLFs in the \emph{fossil-record} scenario trace the shape of the proto-MW/LG SMFs/UVLFs between  $10^{5} \, M_\odot \lesssim M_\star \lesssim 10^{6.5} \, M_\odot$ and $-14 \lesssim M_{\mathrm{UV}} \lesssim - 10$, respectively. The \emph{fossil-record} results differ from the \text{all-progenitors} case because the latter approach combines multiple low-mass/faint progenitors into a single more massive system. 
This difference is compounded in the case of disrupted galaxies which are generally more massive and have multiple reionization-progenitors. 
We also compare our results with the differential SMFs/UVLFs of the FIREbox$\mathrm{^{HR}}$ simulation, as a benchmark for the universal SMF/UVLF, and find close agreement in shape. Although the proto-MW/LG slopes are in agreement with FIREbox$\mathrm{^{HR}}$ within our estimated uncertainties, they are in general flatter by $\sim$0.07. This could be due the difference in the mean matter density of the region around the MW/LG and the mean matter density of the universe, but a detailed exploration is needed, which is beyond the scope of this paper.

\textit{\textbf{The low-mass-end/faint-end slope is more robustly constrained when including  disrupted galaxies.}} We estimate the low-mass-end/faint-end slope of the reconstructed SMFs/UVLFs using a Bayesian approach, to test how accurately and precisely we can recover the proto-MW/LG and the universal SMF/UVLF slopes. In the \textit{all-progenitors} scenario, adding disrupted galaxies helps reproduce the low-mass/faint-end slope of the proto-MW/LG SMF/UVLF more accurately and more precisely, and these are also representative of the global-average low-mass/faint-end slopes. Sample III reproduces the proto-MW/LG slope most accurately with an offset of $\approx$0.04, followed by Samples II ($\approx$0.12) and I ($\approx$0.26). Furthermore, when we use only the intact galaxies, there is a large halo-to-halo scatter due to the varying accretion histories across different halos, even though the median values agree well with that from the proto-MW/LG. Sample III reduces the scatter of Sample I by $\sim$40\%, while Sample II reduces it by $\sim$20\%. Including the disrupted galaxies improves our inferences of the median slopes and also reduces the halo-to-halo scatter. 
In the \emph{fossil-record} scenario, we find similar results when we select high-$z$ progenitors with $M_\star \geq 10^{5} \, M_\odot$ and $M_{\mathrm{UV}} \leq -10$. However, even though the halo-to-halo scatter decreases, we can get a biased estimate of the slope from the fossil record depending on our choice of both the present-day and high-$z$ galaxy samples. The estimated slopes vary depending on the stellar mass range of the galaxies selected, both at the present-day and during the EoR, even in the case of using only intact galaxies. In the future, it may be possible to augment the observed SFHs with average merger histories from cosmological simulations, providing yet more accurate and precise estimates of the low-mass-end/faint-end slope. Nevertheless, including the streams and phase-mixed galaxies improves the statistical constraints on the proto-MW/LG SMFs/UVLFs, reducing the halo-to-halo scatter ($\sim$0.13 in case of proto-MW/LG) from $\sim$0.24 (Sample I) to $\sim$0.20 (Sample II) and $\sim$0.15 (Sample III). A similar trend holds when we separate into MW-like and LG-like environment. This improvement is therefore crucial, since we have only one MW and LG with which to apply the near-far technique.

\textit{\textbf{Adding disrupted galaxies implies a larger contribution of bright star-forming galaxies to the total cosmic UV density in our simulations.}} 
Using the reconstructed UVLFs in the \emph{fossil record} scenario, we compute the cosmic UV density at $z = 7$ and calculate the contribution of different progenitor galaxies to the total universal cosmic UV density. 
We recover only a small fraction of the UV luminosity density budget when including only intact galaxy progenitors, and especially under-predict the amount of UV radiation from galaxies brighter than $M_{\mathrm{UV}} \lesssim -15$. 
Adding disrupted galaxies improves the recovery at these luminosities, and reduces the halo-to-halo scatter in the recovered UV luminosity density. The improved reconstructions at JWST-accessible luminosities with $M_{\mathrm{UV}} \lesssim -15$ will allow more robust comparisons with direct UV luminosity density estimates from JWST UVLFs. 

\textit{\textbf{Applying the near-far technique to streams is feasible with upcoming surveys.}} We estimate that 58\% of our simulated stellar streams, down to $10^6 M_\odot$ in stellar mass, would be detectable using the LSST: 
$\sim 29\%$ of streams with detectable tidal tails, and the others without a prominent tail like intact low-mass galaxies. Roman will enable searches for streams in external galaxies; in this case, we still expect to detect a significant number of streams, but a larger fraction will be classified as intact low-mass galaxies. These preliminary results illustrate the general promise of including stream detections in the near-far technique. 
    

\begin{acknowledgments}
We would like to thank the reviewer for their insightful and valuable feedback which enhanced the clarity of the manuscript.
AK would like to thank Zachary Langford for valuable discussions on Bayesian modeling. We would also like to thank Arpit Arora, Guochao Sun, and Nora Shipp for insightful suggestions, discussions, and comments that have improved the manuscript. We acknowledge use of the \textsf{marmalade} cluster at the University of Pennsylvania, on which the analysis of this work was done.
AK and RES acknowledge support from NSF grants AST-2007232 and AST-2307787. RES further acknowledges support from NASA awards 22-ROMAN22-0055, 22-ROMAN22-0011, and 22-ROMAN22-0013, Simons Foundation grant 1018462, and Sloan Foundation award FG-2023-20669.
AK and AL acknowledge support from the Charles E. Kaufman foundation through grant KA2022-129518.
MBK acknowledges support from NSF grants AST-1910346, AST-2108962, and AST-2408247; NASA grant 80NSSC22K0827; HST-GO-16686, HST-AR-17028, HST-AR-17043, JWST-GO-03788, and JWST-AR-06278 from the Space Telescope Science Institute, which is operated by AURA, Inc., under NASA contract NAS5-26555; and from the Samuel T. and Fern Yanagisawa Regents Professorship in Astronomy at UT Austin.
\end{acknowledgments}

\vspace{5mm}
\software{astropy \citep{Astropy_2013, Astropy_2018, Astropy_2022},
matplotlib \citep{Matplotlib_2007},
IPython \citep{IPython_2007},
NumPy \citep{Numpy_2011, Numpy_2020}, 
SciPy \citep{Scipy_2020},
GizmoAnalysis \citep{GizmoAnalysis_2020},
HaloAnalysis \citep{HaloAnalysis_2020, Wetzel_etal_2016},
PyMC \citep{PyMC},
healpy \citep{Healpy_2019},
scikit-learn \citep{Scikit-learn_2011},
NASA's Astrophysics Data System (ADS),
arXiv
}

\appendix

\restartappendixnumbering

\section{Cumulative Stellar Mass Functions and UV Luminosity Functions}
\label{append_sec:cumul_SMF_UVLF}

\begin{figure*}[ht!]
\centering
\includegraphics[width=\textwidth]{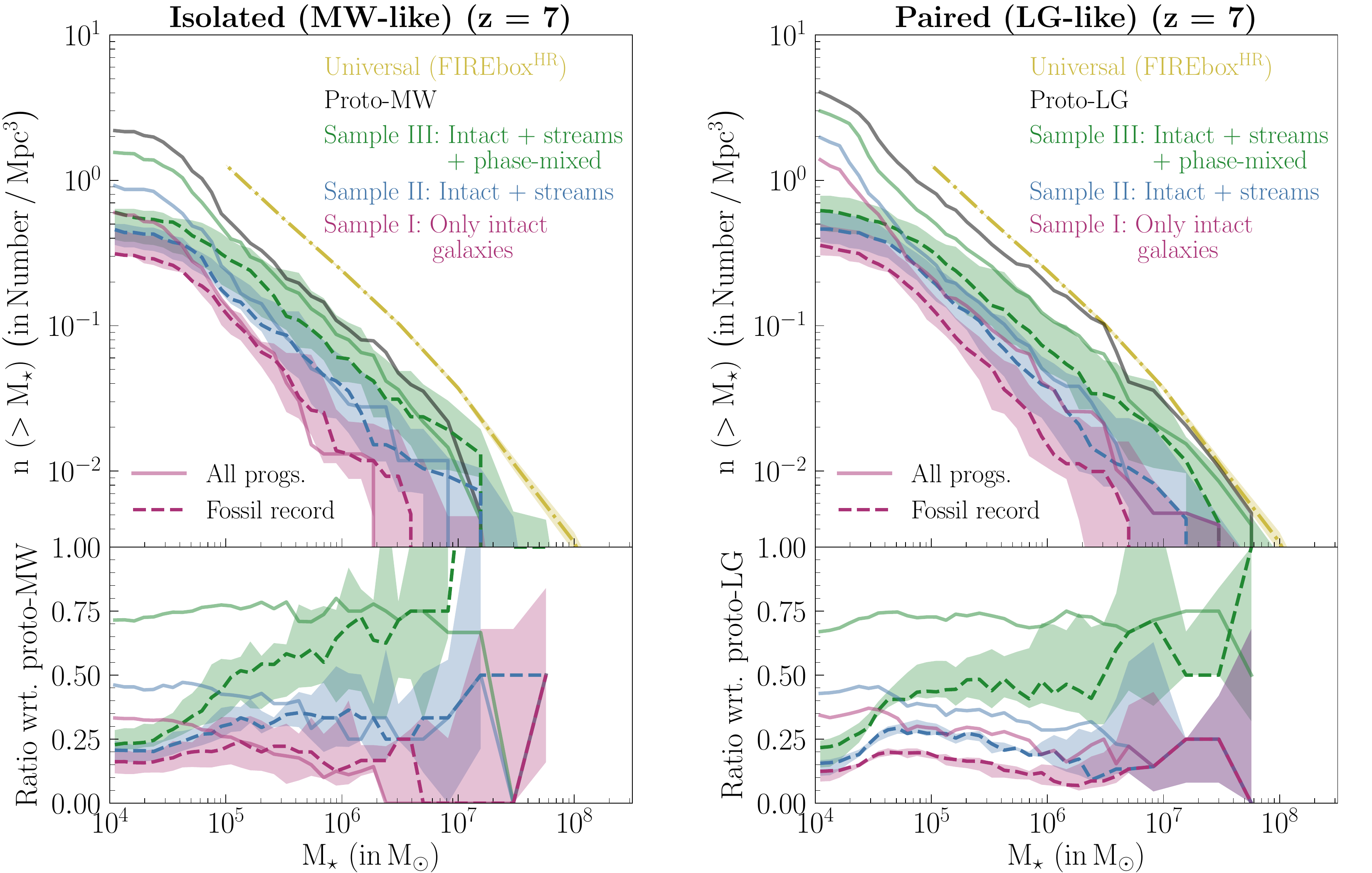}
\caption{\textbf{Cumulative SMFs at $z = 7$}. Same as Figure \ref{fig:7}, but for the cumulative version of the SMFs. The key takeaways are similar to those from Figure \ref{fig:7}.} 
\label{fig:A1}
\end{figure*}

\begin{figure*}[ht!]
\centering
\includegraphics[width=\textwidth]{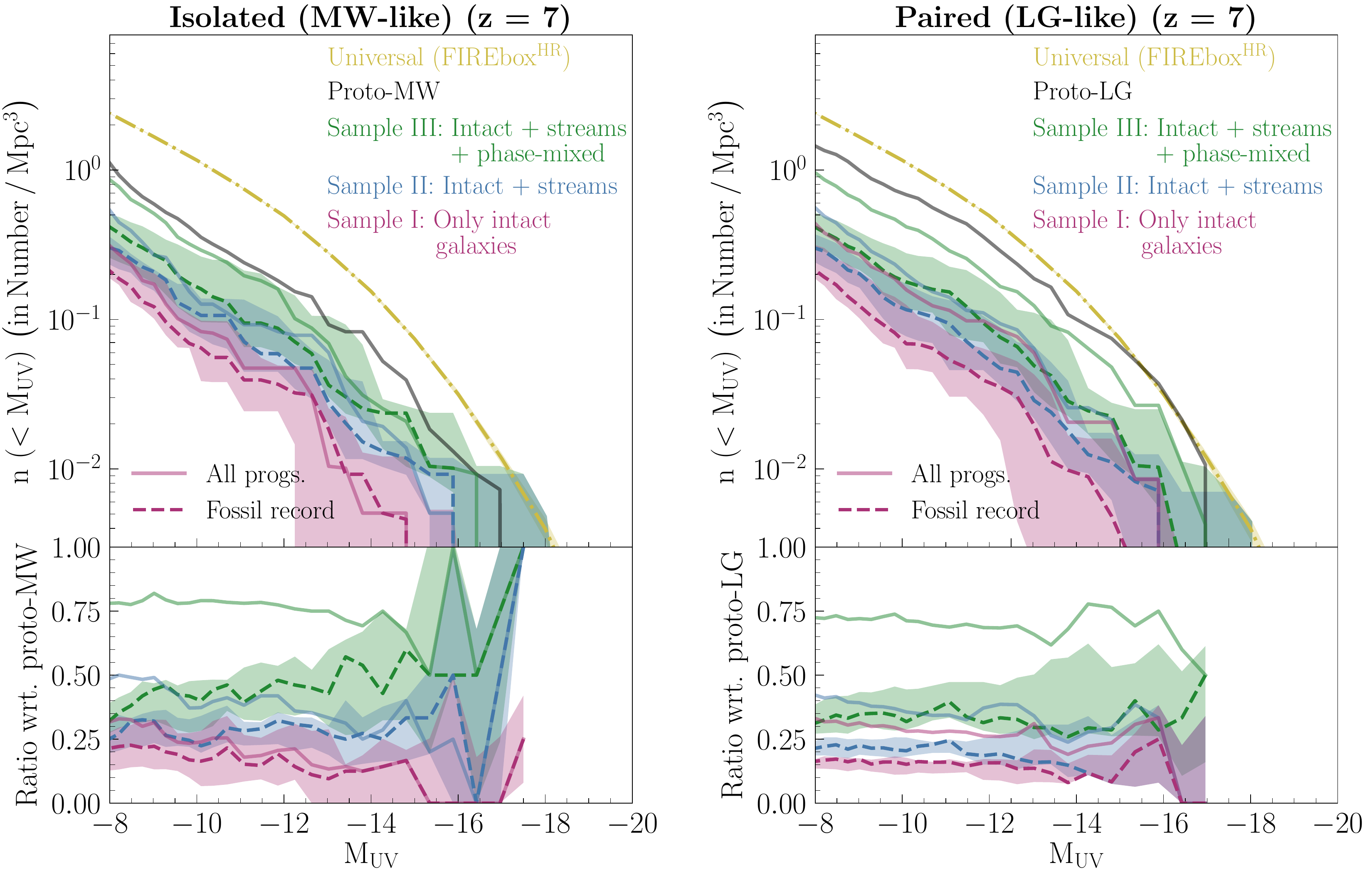}
\caption{\textbf{Cumulative UVLFs at $z = 7$}. Same as Figure \ref{fig:8}, but for the cumulative version of the UVLFs. The key takeaways are similar to those from Figure \ref{fig:8}.} 
\label{fig:A2}
\end{figure*}

Figures \ref{fig:A1} and \ref{fig:A2} show the cumulative versions of the SMFs and the UVLFs at $z = 7$, respectively. Similar to Figures \ref{fig:7} and \ref{fig:8}, we show the median SMFs/UVLFs in the \textit{all progenitors} scenario (solid curves), and in the \emph{fossil record} scenario (dashed curves). The shaded regions show the halo-to-halo scatter at 68\% confidence. The conclusions are similar to what we found from the differential functions. Adding the disrupted galaxies improves the normalization and the shape, particularly in the \textit{all progenitors} scenario. The cumulative SMFs in the `all progenitors' scenario flattens near 10$^4$ $M_\odot$ due to the mass resolution limits of our simulations. This is more prominent in the isolated suite, since their mass resolution is two times coarser than in the paired simulations. In the \emph{fossil record} scenario, the flattening in the cumulative SMFs is mainly due to combining multiple progenitors into a single one, which in turn reduces the number of low-mass galaxies. The flattening effect is less apparent for the cumulative UVLFs because we avoid the extreme
faint end of the luminosity functions which are susceptible to numerical resolution artifacts.

\section{Slope estimates from Bayesian modeling}
\label{append_sec:bayes_SMF_UVLF}

In section \ref{subsec:low-mass-/faint-end_slopes}, we use Bayesian models of Schechter functions to estimate the low-mass/faint-end slopes of the SMFs/UVLFs. This approach avoids binning, which is important because otherwise the slope estimates depend somewhat on the choice of binning scheme. The Bayesian approach described here produces more robust slope determinations. 

Using progenitor galaxies with $M_\star \geq 10^5 \, M_\odot$ and with $M_{\mathrm{UV}} \leq -10$, we estimate the posterior distributions for the $\alpha$ parameter, assuming Schechter functions of the form:

\begin{equation}
    \Phi(\log M) = 10^{\left(\alpha+1\right)\left(\log M - \log M^\star\right)} \, \exp\left(-10^{\log M - \log M^\star}\right) \ln 10
\label{eqn:B1}
\end{equation}

\begin{equation}
    \Phi(M_{\mathrm{UV}}) = \frac{\ln 10}{25} \,\, 10^{0.4 \left(\alpha+1\right)\left(M^\star_{UV} - M_{\mathrm{UV}}\right)} \, \exp\left(-10^{0.4 \left(\alpha+1\right)\left(M^\star_{UV} - M_{\mathrm{UV}}\right)}\right)
\label{eqn:B2}
\end{equation}

where $M$ is the stellar mass, $M_{\mathrm{UV}}$ is the rest-frame UV magnitude, $M^\star$ and $M^\star_{UV}$ are the characteristic stellar mass and UV magnitude, respectively. These describe the shape of the SMFs and UVLFs, while the normalization is set according to:

\begin{equation}
    \Phi_{norm}(\log M) = \frac{\Phi(\log M)}{\displaystyle\int\limits_{\log M_{\mathrm{min}}}^{\log M_{\mathrm{max}}} d(\log M) \, \Phi(\log M)},
\label{eqn:B3}
\end{equation}
and:
\begin{equation}
    \Phi_{norm}(M_{\mathrm{UV}}) = \frac{\Phi(M_{\mathrm{UV}})}{\displaystyle\int\limits_{M_{\mathrm{UV}, \, \mathrm{bright}}}^{M_{\mathrm{UV}, \, \mathrm{faint}}} dM_{\mathrm{UV}} \, \Phi(M_{\mathrm{UV}})}.
\label{eqn:B4}
\end{equation}

\begin{figure*}[ht!]
  \gridline{
    \fig{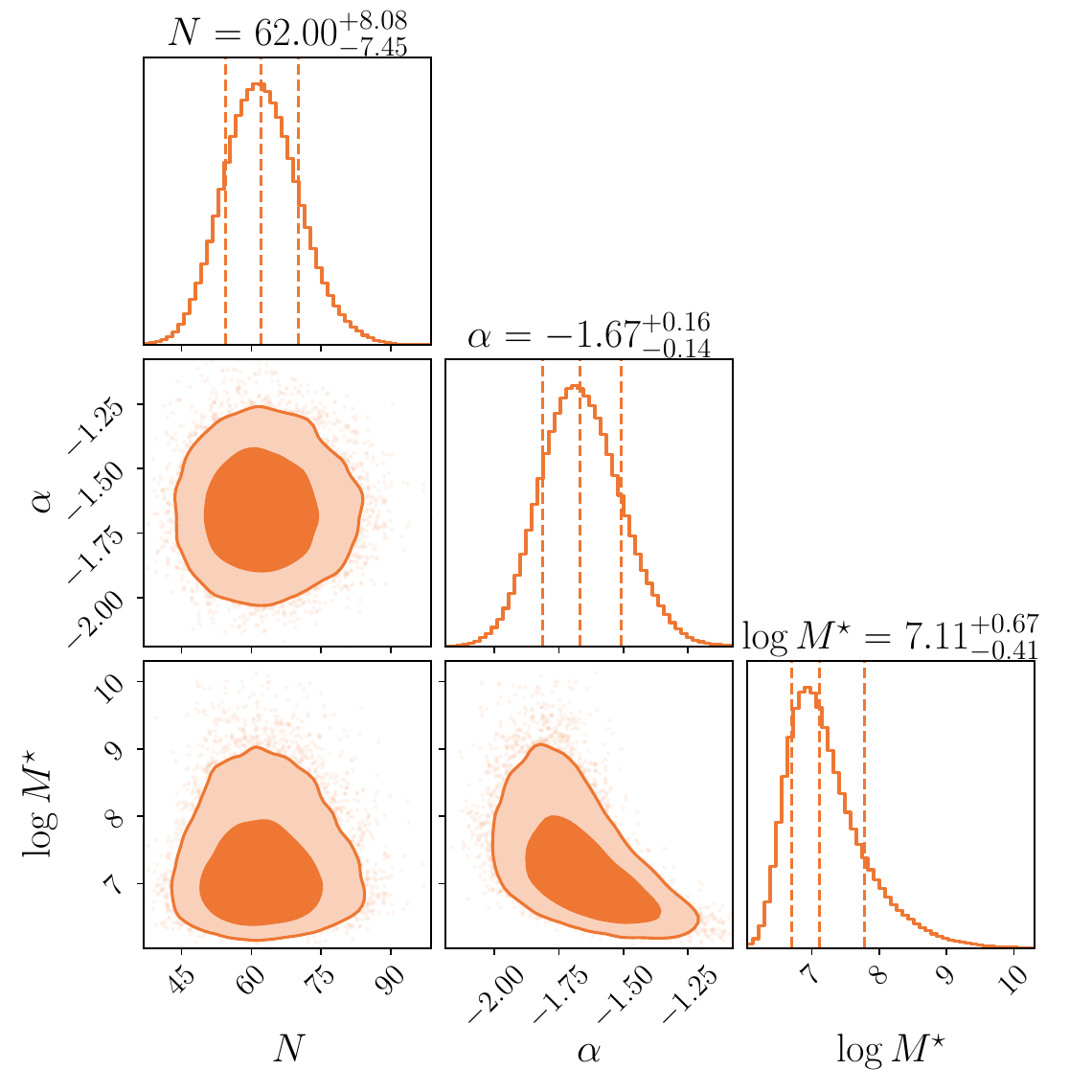}{0.5\textwidth}{}\label{fig:B3a}
    \fig{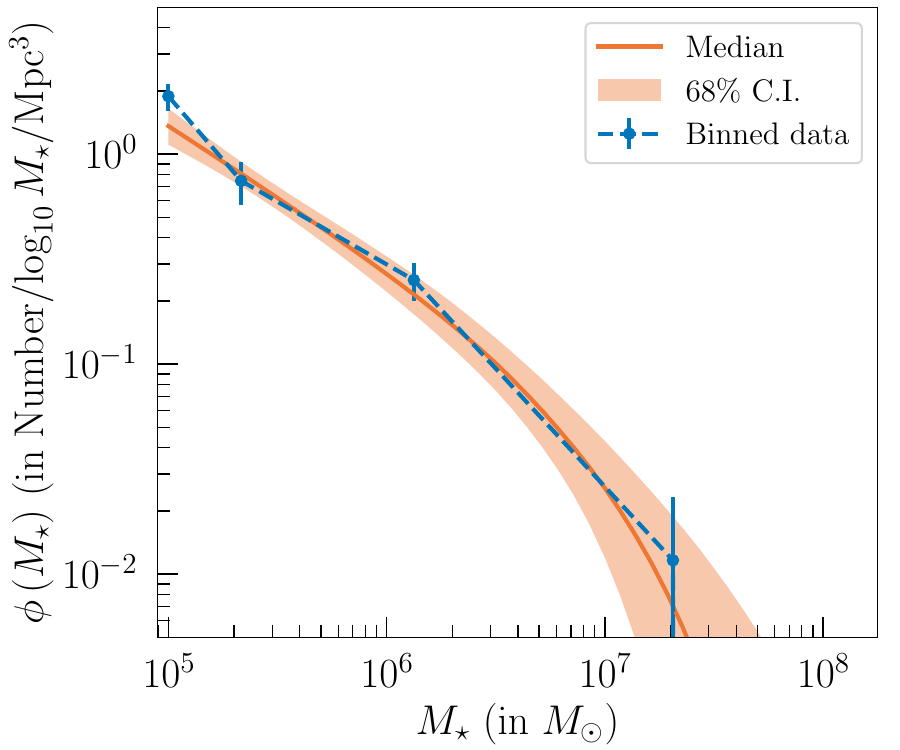}{0.45\textwidth}{}\label{fig:B3b}
  }
  \caption{\textbf{Bayesian modeling of Schechter stellar mass functions for the proto-MW of the \textit{m12i} halo at $z = 7$}. \textit{Left} - The posterior distributions of our parameters $\alpha$, $N$, and $\log M{^\star}$, and their covariances showing 1$\sigma$ and 2$\sigma$ contour intervals. The dotted lines denote the median and 68\% CI of the distributions. \textit{Right} - The blue points denote the binned SMF at $z = 7$, while the orange solid line and shaded regions give the median and 68\% CI among the samples of Schechter stellar mass functions drawn from the parameter posteriors. 
   The binned SMF is well represented by our Schechter function models.}
  \label{fig:B3}
\end{figure*}

Our results are insensitive to the high-mass/bright-end limit of the integrations as long as our highest-mass/brightest progenitor galaxy lies within this limit. The low-mass/faint-end limit does affect our results significantly, as discussed in \ref{subsec:low-mass-/faint-end_slopes}. We count the number of progenitors $n$ in a given sample to estimate the expected number $N$ using a Poisson distribution:
\begin{equation}
    P (n|N) = \frac{N^n \exp(-N)}{n!}.
\label{eqn:B5}
\end{equation}
We incorporate this as a multiplicative factor in our likelihood function, accounting for the Poisson scatter in the number of progenitor galaxies. 
Hence, our log-likelihood function can be written as,

\begin{equation}
    \ln \mathcal{L} = \sum\, \log \, (\Phi_{norm}) + n\log N - N - \log (n!)
    \label{eqn:B6}
\end{equation}

for both the SMFs and the UVLFs.

We adopt truncated normal distributions for the priors on each parameter. For the SMFs we assume:

\begin{align}
    \alpha &\sim \psi(\mu = -1.8, \, \sigma = 0.4, \, a = -3.0, \, b = -1.0) \nonumber \\
    \log M^{\star} &\sim \psi(\mu = 7.0, \, \sigma = 1.0, \, a = 4.0, \, b = 11.0) \nonumber \\
    N &\sim \psi(\mu = 100, \, \sigma = 100, \, a = 1, \, b = 2000),
    \label{eqn:B7}
\end{align}

while for the UVLFs we use:

\begin{align}
    \alpha &\sim \psi(\mu = -1.5, \, \sigma = 0.4, \, a = -3.0, \, b = -1.0) \nonumber \\
    M_{\mathrm{UV}}^{\star} &\sim \psi(\mu = -15.0, \, \sigma = 2.0, \, a = -25.0, \, b = -2.0) \nonumber \\
    N &\sim \psi(\mu = 100, \, \sigma = 100, \, a = 1, \, b = 2000).
    \label{eqn:B8}
\end{align}
Here $\mu$, $\sigma$, $a$, and $b$ denote the mean, standard deviation, lower and upper limits of the truncated normal distributions, respectively. These priors are partially motivated by values obtained in literature and also allow reasonably large range to explore the parameter space. 
Using the \texttt{PyMC} package, we then estimate the posterior distributions from the above priors and likelihood function.

To test whether our method can reproduce the binned SMF, we generate samples of Schechter functions $\Phi(M)$ using the
parameter posteriors. We determine the characteristic normalization, $\Phi{^\star}$, of the Schechter functions according to:
\begin{equation}
    \Phi{^\star} = \frac{N}{V \times \displaystyle\int\limits_{\log M_{\mathrm{min}}}^{\log M_{\mathrm{max}}} d(\log M) \, \Phi(\log M)},
    \label{eqn:B9}
\end{equation}
where $V$ is the convex hull volume used in our calculations. Finally, the normalization from
equation \ref{eqn:B9} is multiplied by the Schechter function shape of equation \ref{eqn:B1} to return a Schechter stellar mass function sample. 
We employ similar technique to generate samples from the Schechter UV luminosity function.  

The left panel of Figure \ref{fig:B3} shows an example corner plot, illustrating the posterior distributions from our analysis. 
The example distributions are for the proto-MW SMF in the \texttt{m12i} halo at $z = 7$. The $\alpha$ and $N$ posteriors appear Gaussian, while that of $\log M{^\star}$ deviates from Gaussianity. The expected number of progenitor galaxies $N$ is also in close agreement with the actual number of progenitor galaxies for the \texttt{m12i} proto-MW with $M_\star \geq 10^{5} M_\odot$. The right panel of Figure \ref{fig:B3} shows how well our method reproduces the SMF in this case, where the solid orange line gives the median of our Schechter stellar mass function samples, while the shaded region denotes the 68\% CI. The blue points show the binned SMF, with Poisson error bars. The Schechter function models agree with the binned measurements, as expected. 

To derive the slopes for the universal SMFs/UVLFs from FIREBox$^{\mathrm{HR}}$, we use a normalized Schechter function with parameters $\alpha$, $\Phi^{\star}$, and $M^{\star}$ or $M^{\star}_{\mathrm{UV}}$, as our likelihood function and fit to the binned SMFs/UVLFs. This is different than what we do for our core suite as discussed above, but we also get posteriors of the parameters using this Bayesian approach.

\section{Bootstrapping Methodology}
\label{append_sec:bootstrap}

\begin{figure*}[ht!]
\centering
\includegraphics[width=\textwidth]{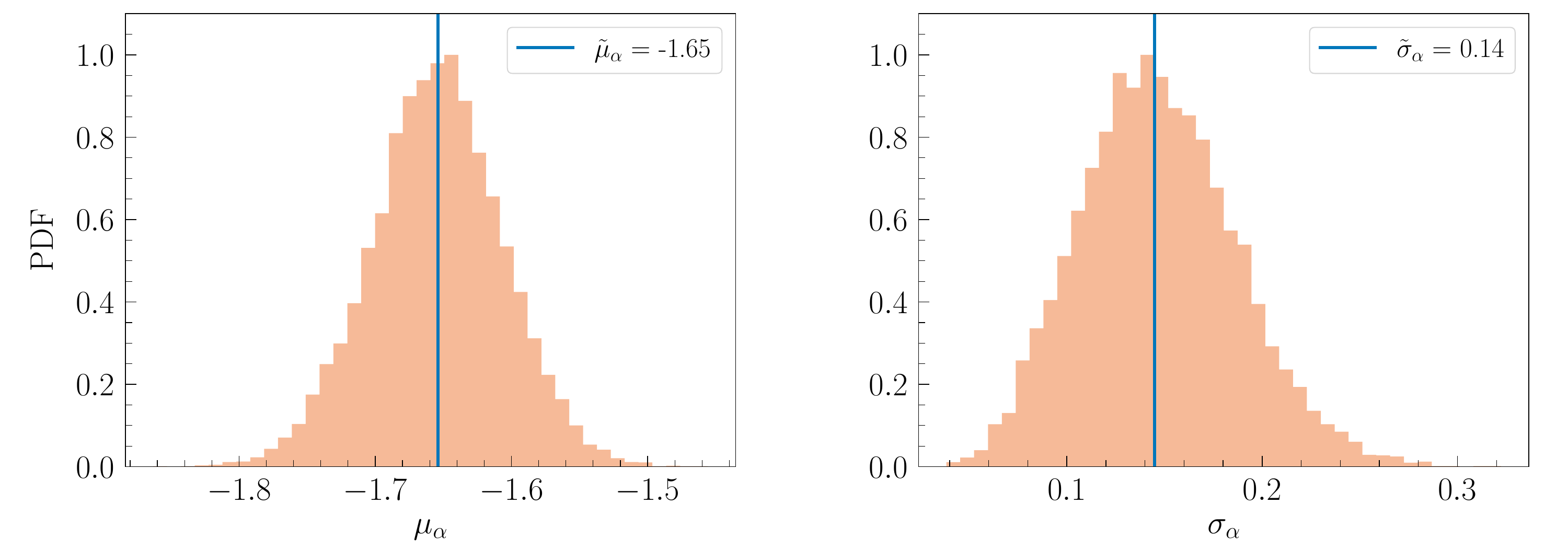}
\caption{\textbf{Distribution of $\mu_\alpha$ and $\sigma_\alpha$ using bootstrapping}. The histograms show the probability density function of $\mu_\alpha$ (\textit{left}) and $\sigma_\alpha$ (\textit{right}) obtained from bootstrapping with replacement for the proto-MW/LG Schechter mass function. The vertical lines denote the median values of these respective distributions which we use as our central estimate and halo-to-halo scatter of the low-mass/faint-end slope respectively.}
\label{fig:C4}
\end{figure*}

In section \ref{subsec:low-mass-/faint-end_slopes}, we used bootstrapping to compute a central estimate of the low-mass/faint-end slope and the scatter across the 10 simulated halos. First, we use the mean and standard deviation of the posterior distribution of the $\alpha$ parameter to draw samples from a Gaussian distribution with this mean and standard deviation. The assumption of Gaussianity is supported by the posterior distribution of $\alpha$ shown in the left panel of Figure \ref{fig:B3}. Next, we use bootstrapping with replacement to select 10 halos at random such that we get 10 values of $\alpha$, and calculate their mean ($\mu_\alpha$) and standard deviation ($\sigma_\alpha$). Then we repeat these two steps 10000 times to get samples of $\mu_\alpha$ and $\sigma_\alpha$. Finally, we compute the median values of these samples, $\tilde{\mu}_\alpha$ and $\tilde{\sigma}_\alpha$, and define them as our central estimate of the low-mass/faint-end slope and the associated halo-to-halo scatter across the 10 halos. We report these values in Table \ref{tab:slopes}. Figure \ref{fig:C4} shows the distribution of our $\mu_\alpha$ and $\sigma_\alpha$ samples for the proto-MW/LG Schechter mass function, with the vertical blue line indicating the median values $\tilde{\mu}_\alpha$ and $\tilde{\sigma}_\alpha$, respectively.

\section{UV Luminosity - Stellar Mass Relation}
\label{append_sec:UV-M_star_relation}

To investigate the shallower slope of the faint-end part of the UVLFs compared to that of the low-mass-end of the SMFs, we first analyze the relation between the UV luminosities and the stellar masses of the progenitor galaxies. We take all of the luminous galaxies from our halo catalogs at $z = 7$, and combine the results across all of our simulations. We then bin in stellar mass and compute the median and 68\% CI in each bin. To fit the relation, we use a power-law of the form:-
\begin{equation}
    L_{\rm UV} = L_{\rm 0} (M_\star/M_{\rm 0})^{\beta}
    \label{eqn:D1}
\end{equation}. 

Our best-fit power-law index is $\beta = 1.22$. From the differential SMF, $\Phi(M_\star)$, we can derive the differential UVLF, $\Phi(M_{\mathrm{UV}})$, using the following equations:-
\begin{equation}
    \Phi(L_{\rm UV}) = \Phi(M_\star)\frac{dM_\star}{dL_{\rm UV}}, 
    \label{eqn:D2}
\end{equation}
and
\begin{equation}
    \Phi(M_{\mathrm{UV}}) = \Phi(L_{\rm UV})\frac{dL_{\rm UV}}{dM_{\mathrm{UV}}},
    \label{eqn:D3}
\end{equation}
where $M_{\mathrm{\rm UV}} \propto -2.5\mathrm{log_{10}}L_{\rm UV}$.

Since the Schechter mass function $\Phi(M_\star)$ follows a power-law of the form $\Phi(M_\star) \propto M_\star^{\alpha_{M_\star}}$ at the low-mass-end, we use the $L_{\rm UV}-M_\star$ relation to derive the form of $\Phi(L_{\rm UV})$. Using equations \ref{eqn:D1}, \ref{eqn:D2}, and \ref{eqn:D3}, we find that $\Phi(M_{\mathrm{UV}})$ also follows a power-law of the form:-
\begin{equation}
    \Phi(M_{\mathrm{UV}}) \propto 10^{-0.4 M_{\mathrm{UV}} (1 + \alpha_{M_\star})/\beta}
\end{equation}.

\begin{figure*}[ht!]
\centering
\includegraphics[width=0.65\textwidth]{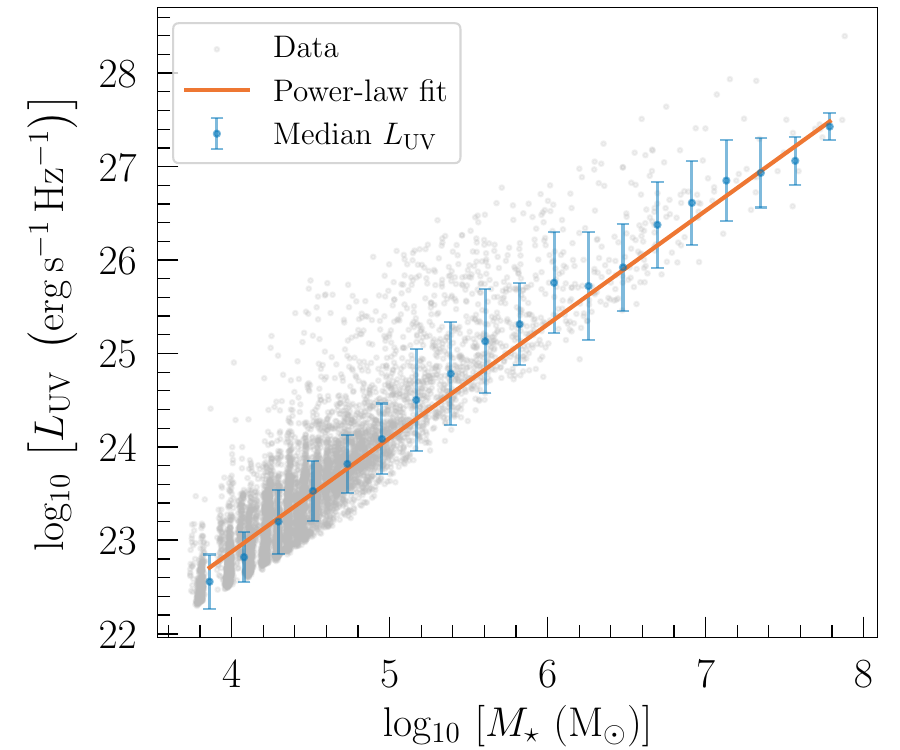}
\caption{\textbf{Superlinear dependence of UV luminosity on the stellar mass at $z = 7$}. 
The grey points show the UV luminosity and stellar mass of each simulated luminous halo at $z=7$.
We bin the points according to their stellar mass. The blue points shows the median values for the UV luminosities in each bin, with the errorbars showing the standard deviations in the respective bins. The orange line shows the best-fit power-law relation between $L_{UV}$ and $M_\star$. A best-fit power-law index of 1.22 shows a superlinear dependence of the UV luminosity on stellar mass, which in turns leads to a shallower faint-end slope for the UVLF compared to the low-mass-end of the SMF.} 
\label{fig:D5}
\end{figure*}

The Schechter function at the fain-end is $\Phi(M_{\mathrm{UV}})$ $\propto 10^{-0.4 M_{\mathrm{UV}} (1 + \alpha_{\rm UV})}$, and so we can write $\alpha_{\rm UV} = \frac{1 + \alpha_{M_\star}}{\beta} - 1$. Using the low-mass-end slope of the proto-MW/LG differential SMF from table \ref{tab:slopes} ($\alpha_{M_\star} =$ -1.65), we get $\alpha_{M_{\mathrm{UV}}}$ = -1.53 for $\Phi(M_{\mathrm{UV}})$, which is a bit shallower, but still differs from our main results (see table \ref{tab:slopes}). Hence, the shallower faint-end slopes of the UVLFs compared that of the low-mass-end of the SMF can be partially attributed to the superlinear dependence of the UV luminosity on the stellar mass. This is because the UV radiation is mostly generated by young stars and the smaller galaxies are not as efficient at forming them, which leads to a suppression at the faint-end. We show the relation and its corresponding linear fit in Figure \ref{fig:D5}. The gray points show the entire set of luminous galaxies from our halo catalogs, the blue points with errorbars show the median $L_{\mathrm{UV}}$ and 68\% CI in each stellar mass bin, and the orange line is the linear fit to the relation.

\bibliography{near_far_disrupted}{}
\bibliographystyle{aasjournal}



\end{document}